\documentclass[oneside]{ieeeoj}

\usepackage{cite}
\usepackage{amsmath,amssymb,amsfonts}
\usepackage{algorithmic}
\usepackage{graphicx,color}
\usepackage{textcomp}
\def\BibTeX{{\rm B\kern-.05em{\sc i\kern-.025em b}\kern-.08em
    T\kern-.1667em\lower.7ex\hbox{E}\kern-.125emX}}
\AtBeginDocument{\definecolor{ojcolor}{cmyk}{0.93,0.59,0.15,0.02}}
\def\OJlogo{\vspace{-12pt}\includegraphics[height=24pt]{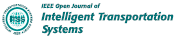}}

\usepackage{geometry}
\usepackage{cite}
\usepackage{amsmath,amssymb,amsfonts}
\usepackage{algorithmic}
\usepackage{lipsum}
\usepackage{subcaption}
\usepackage{graphicx}
\usepackage{textcomp}
\usepackage{xcolor}
\usepackage{comment}
\usepackage{soul}
\usepackage{url}
\usepackage[nolist]{acronym}
\usepackage{multirow} 
\usepackage{array}

\newcommand{\claudia}[1]{\textcolor{magenta}{#1}}

\newcommand{\red}[1] {\textcolor[rgb]{1.0,0.0,0.0}{{#1}}}

\begin{acronym} 
\acro{3GPP}{Third Generation Partnership Project}
\acro{2D}{two-dimensional}
\acro{2G}{second generation}
\acro{3G}{third generation}
\acro{4G}{fourth generation}
\acro{5G}{fifth generation}
\acro{6G}{sixth generation}
\acro{5GAA}{5G Automotive Association}
\acro{5GS}{5G system}
\acro{ACC}{adaptive cruise control}
\acro{AD}{autonomous driving}
\acro{ADAS}{advanced driver assistance systems}
\acro{AECC}{Automotive Edge Computing Consortium}
\acro{AF}{Application Function}
\acro{AGC}{automatic gain control}
\acro{AGV}{automated guided vehicle}
\acro{AI}{artificial intelligence}
\acro{AIaaS}{AI-as-a-service}
\acro{AIFS}{arbitration inter-frame space}
\acro{AMC}{adaptive modulation and coding}
\acro{AoA}{angle of arrival}
\acro{AoI}{age of information}
\acro{API}{application programming interface}
\acro{AS}{application server}
\acro{AD}{autonomous driving}
\acro{AWGN}{additive white Gaussian noise}
\acro{BEP}{beacon error probability}
\acro{BM-SC}{broadcast multicast service center}
\acro{BP}{beacon periodicity}
\acro{B-CSA}{broadcast \ac{CSA}}
\acro{BSM}{basic safety message}
\acro{BSS}{basic service set}
\acro{BS}{base station}
\acro{C-ACC}{cooperative adaptive cruise control}
\acro{C-GLOSA}{cooperative GLOSA}
\acro{C-ITS}{cooperative-intelligent transport systems}
\acro{C-NOMA}{cooperative \ac{NOMA}}
\acro{C-V2X}{cellular-\ac{V2X}}
\acro{C2C-CC}{Car 2 Car Communication Consortium}
\acro{CA}{collision avoidance}
\acro{CACC}{cooperative adaptive cruise control}
\acro{CAM}{cooperative awareness message}
\acro{CAN}{Controller Area Network}
\acro{CAV}{connected and autonomous vehicle}
\acro{CBR}{channel busy ratio}
\acro{CC}{cruise control}
\acro{CCA}{clear channel assessment}
\acro{CCAM}{cooperative, connected and automated mobility}
\acro{CCDF}{complementary cumulative distribution function}
\acro{CDF}{cumulative distribution function}
\acro{CDMA}{code-division multiple access}
\acro{CEPT}{European Conference of Postal and Telecommunications Administrations}
\acro{CoMP}{coordinated multi-point}
\acro{CMM}{co-channel coexistence mitigation methods}
\acro{CN}{core network}
\acro{CP-OFDM}{cyclic prefix orthogonal frequency-division multiplexing}
\acro{CP}{collective perception}
\acro{CPM}{collective perception message}
\acro{CPS}{collective perception service}
\acro{CR}{channel occupancy ratio}
\acro{CRDSA}{contention resolution diversity slotted ALOHA}
\acro{CRLB}{Cram\'er-Rao Lower Bound }
\acro{CSA}{coded-slotted ALOHA}
\acro{CSD}{cyclic shift diversity}
\acro{CSI}{channel state information}
\acro{CSIT}{channel state information at the transmitter}
\acro{CSIR}{channel state information at the receiver}
\acro{CSMA/CA}{carrier sense multiple access with collision avoidance}
\acro{CSMS}{cyber security management system}
\acro{CSSR}{candidate single-subframe resource}
\acro{CTS-To-Self}{Clear-To-Send-To-Self}
\acro{DA}{data age}
\acro{D2D}{device-to-device}
\acro{DC}{duty cycle}
\acro{DCC}{decentralized congestion control}
\acro{DCF}{distributed coordination function}
\acro{DCI}{downlink control
information}
\acro{DCM}{dual carrier modulation}
\acro{DENM}{decentralized environmental notification message}
\acro{DLT}{distributed ledger technology}
\acro{DMRS}{demodulation reference signal}
\acro{DoS}{denial-of-service}
\acro{DDoS}{distributed denial-of-service}
\acro{DOT}{Department of Transportation}
\acro{DRX}{discontinuous reception}
\acro{DSRC}{dedicated short range communication}
\acro{DT}{digital twin}
\acro{DtS}{direct-to-satellite}
\acro{DtS-IoT}{direct-to-satellite Internet of Things}
\acro{EC}{European Commission}
\acro{ECP}{extended cyclic prefix}
\acro{EDCA}{enhanced distributed coordination access}
\acro{EDGE}{Enhanced Data rates for GSM Evolution}
\acro{EED}{eEnd-to-end delay}
\acro{eMBMS}{evolved Multicast Broadcast Multimedia Service}
\acro{eMBB}{enhanced mobile broadband}
\acro{ENI}{ETSI Experiential Networked Intelligence}
\acro{eNodeB}{evolved NodeB}
\acro{EPC}{Evolved Packet Core}
\acro{EPS}{Evolved Packet System}
\acro{ES}{energy signal}
\acro{ETSI}{European Telecommunications Standards Institute}
\acro{eV2X}{enhanced V2X}
\acro{FANET}{flying ad hoc network}
\acro{FCC}{Federal Communications Commission}
\acro{FCD}{floating car data}
\acro{FD}{full-duplex}
\acro{FDA}{feature distribution-aware aggregation}
\acro{FDD}{frequency division duplex}
\acro{FDMA}{frequency division multiple access}
\acro{FEC}{forward error correction}
\acro{FR1}{frequency range 1}
\acro{FR2}{frequency range 2}
\acro{G2G}{glass-to-glass}
\acro{GapRe}{gap sensing and resource re-selection}
\acro{GLOSA}{green light optimal speed advisory}
\acro{GNSS}{global navigation satellite system}
\acro{GPRS}{general packet radio service }
\acro{GPS}{global positioning system}
\acro{GPU}{graphics processing unit}
\acro{HARQ}{hybrid automatic repeat request}
\acro{HD}{high-definition}
\acro{HMI}{human-machine interfaces}
\acro{HPC}{high performance computing}
\acro{HRLLC}{hyperreliable
and low-latency communication}
\acro{HSDPA}{High Speed Downlink Packet Access}
\acro{HSPA}{High Speed Packet Access}
\acro{IBE}{in-band emission}
\acro{ICI}{inter-carrier interference}
\acro{IDMA}{interleave-division multiple access}
\acro{IM}{index modulation}
\acro{IoT}{Internet of Things}
\acro{IP}{Internet Protocol}
\acro{IPG}{inter-packet gap}
\acro{IR}{incremental redundancy}
\acro{IRS}{intelligent reflective surface}
\acro{ISAC}{integrated sensing and communication}
\acro{ISG}{Industry Specification Group}
\acro{ISI}{inter-symbol interference}
\acro{ISM}{industrial, scientific, and medical}
\acro{ISO}{international standardization organization}
\acro{ITU}{International Telecommunication Union}
\acro{ITU-R}{International Telecommunication Union Radiocommunication Sector}
\acro{ITS}{intelligent transport system}
\acro{i.i.d.}{independent identically distributed}
\acro{JSON}{JavaScript object notation}
\acro{KPI}{key performance indicator}
\acro{KRR}{knowledge representation and reasoning}
\acro{LBT}{listen-before-talk}
\acro{LDPC}{low-density parity-check}
\acro{LDS}{low-density spreading}
\acro{LEO}{low Earth orbit}
\acro{LIDAR}{light detection and ranging}
\acro{LLM}{large language model}
\acro{LLR}{log-likelihood ratio}
\acro{LOD}{Linked Open Data}
\acro{LOS}{line-of-sight}
\acro{LoRa}{Long Range}
\acro{LoRa-CSS}{Long Range chirp spread spectrum}
\acro{LoRaWAN}{Long Range wide area network}
\acro{LMF}{Location Management Function}
\acro{LPWAN}{low-power wide-area network}
\acro{LR-FHSS}{long range frequency hopping spread spectrum}
\acro{LTE}{long term evolution}  
\acro{LTE-D2D}{\ac{LTE} with \ac{D2D} communications}  
\acro{LTE-LAA}{LTE license assisted access}
\acro{LTE-V2X}{long-term-evolution-vehicle-to-anything} 
\acro{LTE-V2V}{\ac{LTE}-\ac{V2V}}
\acro{LwM2M}{lightweight machine-to-machine} 
\acro{MAC}{medium access control}
\acro{MAE}{mean absolute error}
\acro{MANET}{mobile ad hoc network}
\acro{MBMS}{Multicast Broadcast Multimedia Service}
\acro{MBMS-GW}{MBMS Gateway}
\acro{MBSFN}{MBMS Single Frequency Network}
\acro{MCE}{Multi-cell Coordination Entity}
\acro{MEC}{multi-access edge computing}
\acro{MCM}{maneuver coordination message}
\acro{MCS}{modulation and coding scheme}
\acro{MIMO}{multiple input multiple output}
\acro{ML}{machine learning}
\acro{mMTC}{massive machine-type communications}
\acro{MPDA}{multi-agent perception domain adaptation}
\acro{MRC}{maximum ratio combining}
\acro{MRD}{maximum reuse distance}
\acro{MUD}{multi-user detection}
\acro{NAV}{network allocation vector}
\acro{NB-IoT}{Narrowband IoT}
\acro{NEF}{Network Exposure Function}
\acro{NGSI-LD}{next generation service interfaces - linked data}
\acro{NGV}{Next Generation V2X}
\acro{NHTSA}{National Highway Traffic Safety Administration}
\acro{NIS2}{network and information systems directive}
\acro{NLOS}{non-line-of-sight}
\acro{NN}{neural network}
\acro{NOMA}{non-orthogonal multiple access}
\acro{NOMA-MCD}{\ac{NOMA}-mixed centralized/distributed}
\acro{NR}{New Radio}
\acro{NR-U}{NR unlicensed}
\acro{NTN}{non-terrestrial network}
\acro{NWDAF}{Network Data Analytics Function}
\acro{OBU}{on-board unit}
\acro{OCB}{outside of the context of a basic service set}
\acro{OEM}{original equipment manufacturer}
\acro{OFDM}{orthogonal frequency-division multiplexing}
\acro{OFDMA}{orthogonal frequency-division multiple access}
\acro{OMA}{orthogonal multiple access}
\acro{OTFS}{orthogonal time frequency space}
\acro{pdf}{probability density function}
\acro{PAPR}{peak to average power ratio}
\acro{PCF}{Policy Control Function}
\acro{PCM}{platoon control message}
\acro{PD}{packet delay}
\acro{PEP}{pairwise error probability} 
\acro{PER}{packet error rate}
\acro{PIAT}{packet inter-arrival time}
\acro{PLS}{physical layer security}
\acro{PPDU}{physical layer protocol data unit}
\acro{PRB}{physical resource block}
\acro{PSCCH}{Physical Sidelink Control Channel}
\acro{PSDU}{physical layer convergence protocolservice data unit}
\acro{PSFCH}{Physical Sidelink Feedback channel}
\acro{PSSCH}{Physical Sidelink Shared channel}
\acro{PHY}{physical}
\acro{PL}{path loss}
\acro{PLMN}{Public Land Mobile Network}
\acro{PPPP}{ProSe Per-Packet Priority}
\acro{PPPR}{ProSe Per-Packet Reliability}
\acro{PQoS}{predictive quality of service}
\acro{ProSe}{Proximity-based Services}
\acro{PRR}{packet reception ratio}
\acro{QAM}{quadrature amplitude modulation}
\acro{QC-LDPC}{quasi-cyclic \ac{LDPC}}
\acro{QoS}{quality of service}
\acro{QPSK}{quadrature phase shift keying}
\acro{RAN}{radio access network}
\acro{RAT}{radio access technology}
\acro{RCS}{radar cross-section}
\acro{RE}{resource element}
\acro{RF}{radio frequency}
\acro{RFID}{radio-frequency identification}
\acro{RIS}{reconfigurable intelligent  surfaces}
\acro{RL}{reinforcement learning}
\acro{RNN}{recurrent neural network}
\acro{RR}{radio resource}
\acro{RRC}{radio resource control}
\acro{RRI}{resource reservation interval}
\acro{RSRP}{reference signal received power}
\acro{RSRQ}{reference signal received quality}
\acro{RSSI}{received signal strength indicator}
\acro{RSU}{road side unit} 
\acro{RTT}{round trip time}
\acro{RTK}{real-time kinematic}
\acro{SAE}{Society of Automotive Engineers}
\acro{SAI}{Service Area Identifier}
\acro{SB-DS}{sensing-based dynamic scheduling}
\acro{SB-SPS}{sensing-based semi-persistent scheduling}
\acro{SC-FDMA}{single carrier frequency division multiple access}
\acro{SC-PTM}{single cell point To multipoint}
\acro{SCS}{subcarrier spacing}
\acro{SCMA}{sparse-code multiple access}
\acro{SDIoV}{software-defined internet of vehicles}
\acro{SFI}{self-interference}
\acro{SFIC}{self-interference cancellation}
\acro{SIFS}{short inter-frame space}
\acro{S-UE}{scheduling UE}
\acro{SCI}{sidelink control information}
\acro{SDG}{sustainable development goals}
\acro{SDN}{software-defined networking}
\acro{SF}{spreading factor}
\acro{SFFT}{symplectic finite Fourier transform}
\acro{SG}{Study Group}
\acro{SHINE}{simulation platform for heterogeneous interworking networks}
\acro{SIC}{self-interference cancellation}
\acro{SINR}{signal-to-interference-plus-noise ratio}
\acro{SNR}{signal-to-noise ratio}
\acro{SRS}{sounding reference signal}
\acro{SSIC}{successive interference cancellation}
\acro{STBC}{space time block codes}
\acro{SV}{smart vehicle}
\acro{TB}{transport block}
\acro{TBC}{time before change}
\acro{TDD}{time-division duplex}
\acro{TDMA}{time-division multiple access}
\acro{ToD}{tele-operated driving}
\acro{TPU}{tensor processing unit}
\acro{TR}{transmission range}
\acro{TS}{technical specification}
\acro{TSG}{Technical Specification Group}
\acro{TTI}{transmission time interval}
\acro{UAV}{uncrewed aerial vehicle}
\acro{UE}{user equipment}
\acro{UD}{update delay}
\acro{UDD}{urban digital development}
\acro{UMTS}{universal mobile telecommunications system}
\acro{URI}{uniform resource identifier} 
\acro{URLLC}{ultra-reliable and ultra-low latency communications}
\acro{USIM}{universal subscriber identity module}
\acro{UTDOA}{uplink time difference of arrival}
\acro{UWAC}{underwater acoustic communications}
\acro{UWB}{ultra-wideband}
\acro{V2C}{vehicle-to-cellular} 
\acro{V2N}{vehicle-to-network}
\acro{V2I}{vehicle-to-infrastructure} 
\acro{V2P}{vehicle-to-pedestrian}
\acro{V2R}{vehicle-to-roadside}
\acro{V2V}{vehicle-to-vehicle} 
\acro{V2X}{vehicle-to-everything} 
\acro{VANET}{vehicular ad hoc network}
\acro{VAM}{VRU awareness message}
\acro{VoI}{value of information}
\acro{VRU}{vulnerable road user}
\acro{WAVE}{wireless access in vehicular environment}
\acro{WBSP}{wireless blind spot probability}
\acro{WLAN}{wireless local area network}
\acro{WPAN}{wireless personal area network}
\acro{XAI}{explainable artificial intelligence}
\acro{XML}{extensible markup language}
\acro{XR}{extended reality}
\end{acronym}

\begin{document}
\receiveddate{XX Month, XXXX}
\reviseddate{XX Month, XXXX}
\accepteddate{XX Month, XXXX}
\publisheddate{XX Month, XXXX}
\currentdate{30 Oct, 2026}
\doiinfo{10.1109/OJITS.2026.3749067}


\title{\huge 6G-Enabled Cooperative Connected and Automated Mobility: Are We There Yet?}

\author{
    Mate Boban$^{1}$,
    Alessandro Bazzi$^{2}$,
    Claudia Campolo$^{3}$,
    Alexey Vinel$^{4}$,
    Mario Castaneda$^{1}$,
    Luca De Nardis$^{5}$,
    Eirini Liotou$^{6}$,
    Jérôme Härri$^{7}$,
    Nikolaos Koursioumpas$^{8}$,
    Lina Magoula$^{8}$,
    Barbara Masini$^{9}$,
    Antonella Molinaro$^{3}$,
    Lei Wan$^{4,10}$, and
    Tommaso Zugno$^{1}$
    \\
    \vspace{1mm}
    {\footnotesize 
    $^{1}$Huawei Technologies Duesseldorf GmbH \quad
    $^{2}$Università di Bologna \quad
    $^{3}$Università Mediterranea di Reggio Calabria \quad
    $^{4}$Karlsruhe Institute of Technology \quad
    $^{5}$DIET Department, Sapienza University of Rome \& WiLab, CNIT \quad
    $^{6}$Dept. of Informatics and Telematics, Harokopio University of Athens\quad
    $^{7}$EURECOM \quad
    $^{8}$Dept. of Informatics and Telecommunications, National and Kapodistrian University of Athens \quad
    $^{9}$National Research Council (CNR) of Italy \& Wilab, CNIT\quad
    $^{10}$XITASO GmbH\\
    }
}
\markboth{Journal,~Vol., No., Date}%
{Boban \MakeLowercase{\textit{et al.}}: 6G-Enabled Cooperative Connected and Automated Mobility: Are We There Yet?}

\corresp{Corresponding author: Mate Boban (e-mail: mate.boban@huawei.com).}

\begin{abstract}

\textcolor{black}{
Connected, cooperative, and automated mobility (CCAM) is a key enabler of future intelligent transportation systems, requiring seamless integration of communication, sensing, computing, and automation technologies. This paper surveys the state of the art in CCAM, focusing on key use cases and performance indicators, enabling technologies, system architectures, and standardization and regulatory developments relevant to large-scale deployment. Particular attention is given to cooperative perception, collective awareness, artificial intelligence and machine learning as fundamental building blocks of future CCAM ecosystems. The paper further identifies major research challenges related to scalability, interoperability, and real-time operation in highly dynamic mobile environments. Finally, it outlines future research directions toward achieving large-scale, safe, and resilient CCAM systems supported by next-generation communication networks.}

\end{abstract}



\maketitle

\section{Introduction}
\label{intro}


The evolution of transportation systems is increasingly driven by the convergence of automation, wireless connectivity, and cooperative intelligence, embodied in the concept of \ac{CCAM}. 

Concerning automated driving, despite significant advances, the current state-of-the-art is mainly focused on Levels 2 to 3 automation, offering \ac{ADAS} and limited \ac{AD} capabilities.
In terms of connectivity, there is a continuous technical progress on the different \ac{V2X} communication technologies~\cite{10488039}, including parts of the evolution of  cellular networks that are currently progressing towards \ac{6G}; however, while there are some limited deployments in different markets, there is no massive scale deployment of V2X systems.  

\textcolor{black}{
This gap between technological advances and large-scale deployment is increasingly recognized by industry, policymakers, and public stakeholders. Reflecting this momentum, the Smart Networks and Services Joint Undertaking (SNS JU) recently published the \emph{5G Strategic Deployment Agenda for Connected and Automated Mobility on Roads,\footnote{\url{https://tinyurl.com/SNSEU5GCAM}}} outlining a roadmap for deploying connected and automated mobility services along major European transport corridors. Building on extensive \ac{5G} corridor trials, the initiative signals a transition from experimental validation toward operational deployment of CCAM services and highlights the need for communication, sensing, and computing capabilities that can support seamless, scalable, and cross-border mobility. These requirements further motivate the exploration of \ac{6G} as a key enabler of future CCAM ecosystems.
}

\begin{figure*}[t]
\centering
\includegraphics[width=\textwidth,draft=false]{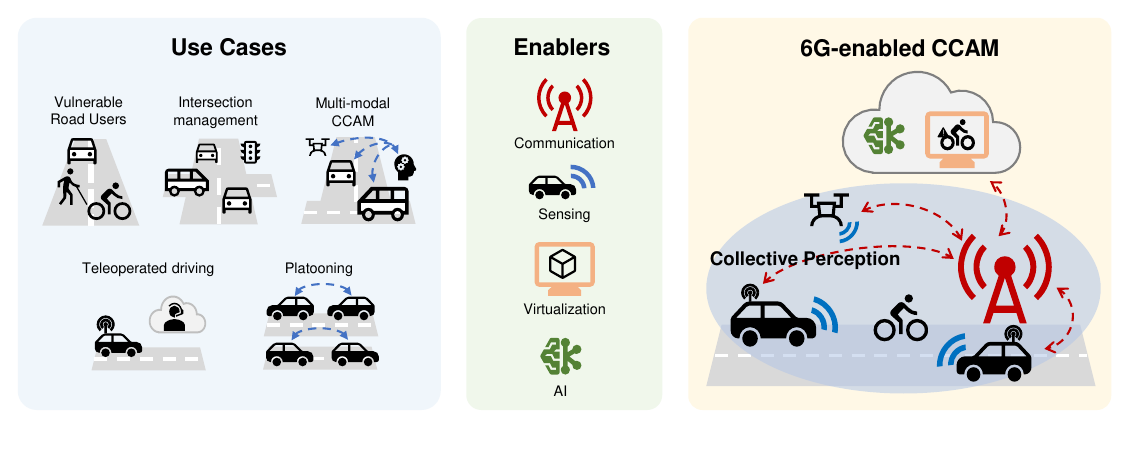}
\caption{6G-enabled CCAM vision. CCAM use cases can be enabled with enhancements that 6G will bring in terms of highly reliable communication and sensing, integration of advanced software architectures (such as \ac{DT}), and deep integration of AI technologies in both the 6G network and services. These will feed into \ac{CP}, which is an essential block for integrating non-cooperative actors, which would in turn enable the full CCAM solution.}
\label{fig:CCAM}
\vskip -0.3cm
\end{figure*}

\subsection{Preliminary Overview}\label{Intro:usecases}

Although CCAM is a cornerstone of further urban digital development, in accordance with sustainable development goals,\footnote{\url{https://sdgs.un.org/goals}}  
current systems struggle with fragmented sensing, limited field of view, and lack of coordination across agents, hindering the scalability and reliability of safety-critical functions. 
A possible game changer at the heart of \ac{CCAM} is represented by  \underline{\textit{\ac{CP}}}~\cite{wan2025systematic}, which refers to the ability of vehicles and infrastructure to share sensor data and contextual information 
in real time. This enables a collective awareness of the traffic environment far beyond what individual entities can achieve alone, and also incorporating the non-connected actors in the network, at least from the perspective of sensing/detecting them.  
\textcolor{black}{
Recent work has further demonstrated the benefits of infrastructure-assisted cooperative perception. For example, Meng and Zhao~\cite{meng2025vi} proposed VI-BEV, a vehicle--infrastructure collaborative perception framework that fuses sensing information from vehicles and \acp{RSU} to improve 3D object detection in bird's-eye-view representations. By leveraging complementary sensing perspectives, the framework extends perception coverage and mitigates occlusions, thereby enhancing environmental awareness beyond the capabilities of standalone perception systems.}

Thanks to the addition of \ac{CP}, \ac{CCAM} 
can address today's most pressing transportation challenges while unlocking a future of \textit{safer, greener, and more integrated and efficient mobility ecosystems}. 
The potential impact of \ac{CCAM} and CP is revolutionary, particularly when applied to the following \underline{\textit{key use case groups}}:  
\textit{(i)} 
\textit{Vulnerable road user (VRU)\acused{VRU} protection}, based on CP-enabled detection and localization of non-connected users such as pedestrians and cyclists~\cite{10814969,zoghlami20235g}. Shared perception data from vehicles or infrastructure allows for early warning and proactive collision avoidance, even in occluded or low-visibility conditions; 
\textit{(ii)} 
\textit{Platooning and chauffeuring}, meaning tightly coupled vehicle groups acting as an unique convoy. This implies reduced inter-vehicle gaps, improved aerodynamics, and fuel savings, thus enhancing efficiency together with safety~\cite{BALADOR2022100460}; 
\textit{(iii)} 
\textit{Intersection Management}, which benefits from real-time communications-enabled exchange of positions, intentions, and priorities between vehicles and infrastructure. CCAM enables in this scenario smoother merging, conflict-free crossing, and reduced delay, which in turn leads to improved traffic flow capacity, energy efficiency, and safety~\cite{10.1145/3407903}; 
\textit{(iv)} 
\textit{Tele-operated driving (ToD)\acused{ToD}}, which refers to the remote operation of vehicles by remote (human or otherwise) drivers located outside the vehicle itself~\cite{gnatzig2013system}. This scenario is particularly relevant in situations where full autonomy is not feasible, 
and enables human oversight and intervention in real time, bridging the gap between the current driver-centric and the future fully automated mobility; 
and \textit{(v)} 
\textit{Multimodal transportation,} where road traffic interacts with public transport, micro-mobility, and logistics services. In this use case, \ac{CCAM} can facilitate safe and efficient coordination, thus supporting smoother integration of various transport modes and optimized infrastructure usage.

\begin{table*}[t]
\centering
\begin{tabular}{p{3.4cm}p{1.5cm}p{2.0cm}p{2.5cm}p{6.2cm}}
\hline
\textbf{First author, year, ref.} & \textbf{Application perspective} & \textbf{Communication perspective} & \textbf{\hbox{Beyond Traditional} Comm. perspective}& \textbf{Main focus} 
 \\
\hline  \hline 
Balador, 2022, \cite{BALADOR2022100460} & $\checkmark$ & ($\checkmark$) & &Platooning  \\
Abdi, 2025, \cite{10814969} & $\checkmark$& ($\checkmark$) & &VRU protection  \\
Zoghlami, 2023, \cite{zoghlami20235g}  & $\checkmark$ &  $\checkmark$ &  &5G-enabled VRU protection \\
Khayatian, 2020, \cite{10.1145/3407903}& $\checkmark$& & &Intersection management  \\

Bagheri, 2021, \cite{Bagheri2021} & & $\checkmark$  &  &5G V2X-enabled cooperative driving\\
Bazzi, 2021, \cite{bazzi2021design} & & $\checkmark$& & 5G NR-V2X specifications and literature overview \\ 
    Casta\~{n}eda Garcia, 2021, \cite{NRV2XR16Tutorial} & ($\checkmark$) & $\checkmark$& & 5G NR-V2X specifications  \\
Pawar, 2024, \cite{10767678} & $\checkmark$& $\checkmark$& &5GAA use cases and novel technologies in 5G  \\
Clancy, 2024, \cite{10488039} & ($\checkmark$) & $\checkmark$ & ($\checkmark$) & Cellular and WiFi based access technologies   \\
Annu, 2024, \cite{10443065} & ($\checkmark$) & $\checkmark$& & 6G sidelink resource allocation schemes   \\
Lang, 2025, \cite{LANG2025110916} & &  & $\checkmark$& Technologies for 6G V2X\\
Noor-A-Rahim, 2022, \cite{9779322} & & & $\checkmark$ & Technologies for 6G V2X \\
Roger, 2024, \cite{10465596} & $\checkmark$ & & ($\checkmark$) & Sustainability, emission-related use cases  \\
\textbf{This work} & \textbf{$\checkmark$}& \textbf{$\checkmark$}& \textbf{$\checkmark$}& \textbf{6G technologies and CCAM}  \\
\hline \hline
\end{tabular}
\caption{Overview of recent survey papers.}
\label{tab:comparison}
\end{table*}

Current wireless technologies may not be sufficient to fully enable CCAM, with CP at its heart and with a requirement to enable the listed challenging use cases. Therefore, work is being done to improve the \underline{\textit{enabling technologies}}. This in part refers to the communication techniques improvements, for example including \ac{FD}, \ac{NOMA}, or \ac{RIS} solutions, and utilizing additional frequency bands (e.g., sub-THz bands). Furthermore, enabling \ac{CCAM} will necessitate going beyond pure communications to include aspects such as \ac{ISAC}, \ac{AI} \textcolor{black}{and \ac{ML}}, \ac{DT}, and network virtualization, which might need to be considered from the beginning of the system design if the promise of CCAM is to be fulfilled.

The purpose of this paper is to analyze all these aspects and provide a starting point for discussing the next steps toward the future of CCAM.




%




\subsection{Related Work: What Did Other Surveys Address?}\label{sec:relatedwork}

Given the recent advances and the potential impact of CCAM, a number of surveys and tutorials have been published on this topic in the recent years, as summarized in Table \ref{tab:comparison}. 
A number of these works look at CCAM from an application level perspective, typically restricting the focus to a specific application and providing a detailed review of the related literature; for example, \ac{VRU} protection is addressed in \cite{10814969,zoghlami20235g}, platooning in  \cite{BALADOR2022100460}, intersection management in \cite{10.1145/3407903}. 
These works are mainly concentrated on the functional aspects and do not discuss the access technologies in detail.

A larger group of review works focus instead on the access technologies, detailing the standards at the physical and \ac{MAC} layers, 
with the majority restricting the attention exclusively to the \ac{C-V2X} ecosystem. In \cite{9217500}, for example, the authors analyze the main aspects of \ac{LTE}-V2X and discuss a number of directions of improvement for the \ac{5G} that was at that time being developed, while works in \cite{Bagheri2021} and \cite{bazzi2021design} discuss the advances introduced by 5G \ac{NR}-V2X and the literature addressing them. 
A deeper discussion of the specifications introduced by the \ac{3GPP} for 5G NR-V2X in Rel-16 is then addressed by \cite{NRV2XR16Tutorial}, which provides details on the physical layer, the resource allocation, and the quality of service management, giving particular emphasis on the sidelink. In \cite{10767678}, the authors provide a more general overview of the main 5G technologies, including those from the perspective of radio (e.g., mmWave) and architecture (e.g., edge computing, network slicing), together with the use cases addressed 
by the \ac{5GAA}. 
With a similar objective, an overview of both WiFi and C-V2X is provided in  \cite{10488039}, associated with the use cases defined by the 3GPP during the design of 5G NR-V2X; a brief discussion of the possible evolutions are provided, but limited to the access layer and network architecture. A projection towards 6G  is then addressed by \cite{10443065}, exclusively focusing on the sidelink and mostly concentrated on the issue of resource allocation, with some early considerations on what is expected in the next generation. 
All these surveys focus on analyzing the signals and communication protocols, while the discussion provided from an application perspective or towards aspects beyond communications, when present, is of minor relevance.

Finally, while the mentioned group of surveys intend to provide insights into access technologies as they are today, other works, such as \cite{LANG2025110916,9779322} specifically address enablers for the next generation; in these cases, the scope is to focus on the enabling technologies of 6G, only briefly touching on the available standards and the CCAM applications that they intend to address. A clearer link to the use cases is provided in \cite{10465596}, where the main focus is on how V2X can improve the sustainability, giving emphasis to the technologies the authors consider more relevant for that target and discussing a number of specific use cases where V2X can make transportation greener. In all these cases, the current status from a research and standardization perspective is overlooked.



In comparison to previous works, the aim of this paper is to provide an overview of a complete ecosystem required for \ac{CCAM}, looking at the available solutions, the main use cases to be addressed, and how the upcoming \ac{6G} technologies could lead to successful deployment of CCAM. The novelty is therefore in the wide perspective, with focus on both the functional operations and the key enablers 
in the 6G realm beyond those designed for a mere connectivity service, encompassing new technologies such as \ac{AI}, \ac{ISAC}, and \acp{DT}, among others. 

\textcolor{black}{While security and privacy are of paramount importance for CCAM, a comprehensive discussion of these topics falls beyond the scope of this survey. Nevertheless, relevant security- and privacy-related considerations are highlighted throughout the paper where appropriate, in particular in Sections~\ref{cp} on \ac{CP} and Section~\ref{keyenab} on key enablers. Readers interested in a more extensive treatment are referred to the recent surveys in \cite{10026338,9108394,10.1145/3558052,11012728,ying2024v2xsecurity}.}

\begin{figure*}[t]
\centering
\includegraphics[width=1.9\columnwidth]{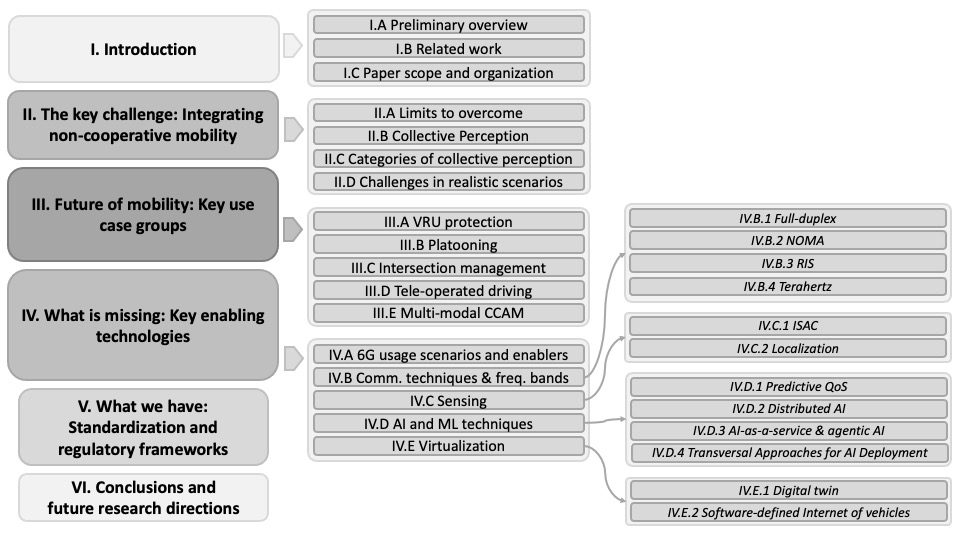}
\caption{\textcolor{black}{Structure of the paper.}}
\label{fig:papermap}
\end{figure*}

\subsection{Paper Scope and Organization}\label{Intro:novelty}

In this paper, we first provide an overview of \ac{CCAM} and its main target scenarios, followed by discussing the current status and ongoing activities related to \ac{CCAM}. Figure~\ref{fig:CCAM} summarizes the use cases and technological enablers discussed in the paper and shows a high level vision of 6G-enabled CCAM which we further elaborate on throughout the paper in the following directions: 
\begin{itemize}
\item We identify a core technical challenge, which is preventing \ac{CCAM} from its ubiquitous deployment, namely the inability to incorporate non-cooperative users into the cooperative transport system, and discuss how CP can be the lever to change the game;
\item We describe the main scenarios targeted by CCAM and, for each of them, detail to what extent the available technologies can support their implementation; 
\item We discuss how key technologies that are not available in current networks, but are expected in 6G, can help enabling CCAM in the coming future, 
analyzing for each of them the open challenges and how hey can be addressed;
\item We summarize the most relevant past and ongoing standardization efforts related to CCAM.
\end{itemize}

The rest of the paper is organized as summarized in Fig.~\ref{fig:papermap}: After the introduction given in this section, Section~\ref{cp} discusses the key \ac{CCAM} problem of integration of non-cooperative users into the connected eco-system. Section~\ref{tarsce} elaborates on the main target use cases for \ac{CCAM} systems; the role of the key enabling technologies, which are foreseen to become a future of standardization within the path of 6G, is dissected in Section~\ref{keyenab}, while Section~\ref{standard} provides an overview of the ongoing standardization efforts for \ac{V2X} communications. Finally, conclusions are reported in Section~\ref{concl}. For an easier reading, a table with the definition of the main acronyms is reported in Table~\ref{tab:acronyms}.

\begin{table}[t]
\centering
\begin{tabular} 
{p{1cm} p{6cm}} \\
\hline
\textbf{Acronym} & \textbf{Definition} \\
\hline \hline 
ADAS & Advanced driver assistance systems \\
AI & Artificial intelligence \\
AIaaS & AI-as-a-service \\
BS & Base station \\
C-ITS & Cooperative-intelligent transport systems \\
C-V2X & Cellular-V2X \\
CACC & Cooperative adaptive cruise control \\
CCAM & Cooperative, connected and automated mobility \\
CP & Collective perception \\
DT & Digital twin \\
FD & Full-duplex \\
FR1 & Frequency range 1 \\
FR2 & Frequency range 2 \\
GNSS & Global navigation satellite system \\
HRLLC & Hyperreliable
and low-latency communication \\
IoT & Internet of Things \\
IRS & Intelligent reflective surface \\
ISAC & Integrated sensing and communication \\
KPI & Key performance indicator \\
MEC & Multi-access edge computing \\
ML & Machine learning \\
NGV & Next Generation V2X \\
NOMA & Non-orthogonal multiple access \\
NR & New Radio \\
OBU & On-board unit \\
PQoS & Predictive quality of service \\
RAT & Radio access technology \\
RIS & Reconfigurable intelligent  surfaces \\
RSU & Road side unit \\ 
SDIoV & Software-defined internet of vehicles \\
SFIC & Self-interference cancellation \\
SDN & Software-defined networking \\
SSIC & Successive interference cancellation \\
ToD & Tele-operated driving \\
UE & User equipment \\
URLLC & Ultra-reliable and ultra-low latency communications \\
V2X & Vehicle-to-everything \\ 
VRU & Vulnerable road user \\
XAI & Explainable AI \\
\hline \hline
\end{tabular}
\caption{Main acronyms used in the paper.}
\label{tab:acronyms}
\end{table}

\section{The Key Challenge: Integrating Non-cooperative Mobility through Collective Perception}
\label{cp}

\subsection{Limits of Cooperative Awareness/Maneuvering}

The central challenge for the ubiquitous deployment of the CCAM has been associated with the fact that many road users in nowadays traffic are not cooperative or not even connected. Indeed, market deployment of V2X technologies is gradual, and the investments made today might not result in short-term benefits. Until now, the attention of V2X stakeholders has been mainly focused on the cooperative awareness service:\footnote{
When looking at the V2X services and messages, in this document we refer to those defined by ETSI; other standardization bodies such as SAE have defined similar concepts with different names.} a periodic broadcasting of \acp{CAM} (in Europe) or similar messages, which enable each vehicle to continuously notify own-status updates to the road users in the immediate vicinity. This improves in principle the vehicular perception beyond visual occlusions and the range limitations of local sensors (e.g., radars, LiDARs, and cameras), but is effective only if the majority if not all of the vehicles are connected and cooperative~\cite{vinel2012}.

Recently, the efforts have been made to design the maneuver coordination service, which, from the CCAM perspective, is the most comprehensive and futuristic V2X-based enabler of road traffic coordination and orchestration~\cite{10575944}.  It is based on the exchange of \acp{MCM} that include not only the current status of an ego vehicle,\footnote{In the context of autonomous driving, an \emph{ego vehicle} is the vehicle under primary interest in a study or a test scenario} but also its intentions (e.g., in the form of the planned trajectory). Unfortunately, the maneuver coordination approach is also rather limited in practice, since it does not allow to integrate in the CCAM whatever is not connected, which means that all obstacles or moving objects for which cars need coordination, should be themselves connected and cooperating.  


This aspect is critical when looking at the CCAM deployment, and makes the recently standardized \ac{CPS}~\cite{ETSI2023} a possible game-changer and a connecting fiber between the use cases and technology enablers depicted in Fig.~\ref{fig:CCAM}. As detailed hereafter, the exchange of \acp{CPM} provide in fact the ability to incorporating also non-cooperative vehicles and non-cooperative objects into the CCAM eco-system.

\subsection{Collective Perception}

The \ac{CPS} is based on the broadcast of \acp{CPM}, which in turn include  information about the objects perceived by the vehicle local sensors~\cite{ETSI2023}. 
When the vehicles share the information not only about their own status, but also about the status of other visible and, perhaps, not connected users or objects, this results in a powerful distributed sensing system which presents a solid immediate basis for the entire CCAM framework. \ac{CP} allows sensor-equipped vehicles and infrastructure to exchange perception data, extending the field-of-view of individual vehicles and thereby helping to overcome visual occlusions and long-range detection limitations. \ac{CP} contributes to a comprehensive situational awareness even in scenarios with low \ac{V2X} penetration rate, making it a strategic enabler for the early deployment and incremental rollout of \ac{CCAM} systems~\cite{AdaptiveFeatureFusion-2023-qiaob}. 

\begin{figure}
    \centering
    \includegraphics[width=0.45\textwidth]{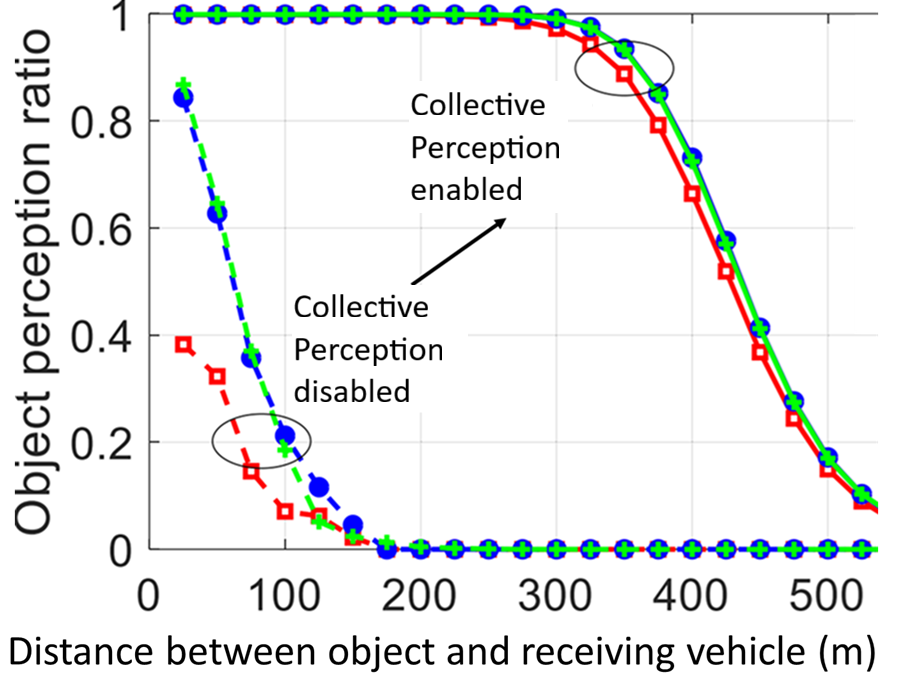}
    \caption{Benefits of collective perception over sensor-only solutions in terms of object perception ratio (figure modified from~\cite{thandavarayan2020cooperative}; red curve: forward sensor; blue curve: 360$^0$ sensor; green: Tesla Autopilot sensors).}
    \label{fig:score}
\end{figure}

Since the transmission of raw sensor data is practically infeasible due to \ac{V2X} network bandwidth limitations, the selection of which information is transmitted 
and the frequency of its inclusion in the \acp{CPM} becomes crucial. Even if a number of different approaches is possible and work is in progress also in standardization, this is in general performed using the concept of \ac{VoI}, which gives a score to the importance of an update for each specific object \cite{lyu2025accuracy}. 

Figure~\ref{fig:score}, reporting results from~\cite{thandavarayan2020cooperative}, highlights the substantial benefits of \ac{CP} by comparing the object perception ratio achieved under different sensor configurations with and without communication support. The results show that \ac{CP} markedly extends the effective perception range, enabling vehicles to detect objects far beyond the limits of their onboard sensors and even in areas outside their own field of view. When \ac{CP} is used, the object perception of 100\% can be obtained up to 250~m, irrespective of which type of sensor is used (red curve: forward sensor; blue curve: 360$^0$ sensor; green: Tesla Autopilot sensors). Sensors without \ac{CP} fail to reach 100\% object perception. This capability also mitigates the impact of occlusions, which strongly degrade direct sensing performance, particularly in denser traffic. Notably, the performance differences between forward-only, 360°, and Tesla sensor setups become minimal once \ac{CP} is enabled, demonstrating that shared sensing information significantly reduces reliance on complex or wide–field-of-view sensor suites. 

\textcolor{black}{
The importance of efficient information sharing in \ac{CP} has also been highlighted by Figueiredo \textit{et al.}~\cite{figueiredo2024enhancing}, who investigated the impact of object data dissemination within the ETSI \ac{CPS}. Their study showed that exchanging object-level perception information can significantly enhance situational awareness beyond the sensing range of individual vehicles, while also introducing communication overhead that must be carefully managed. 
}


Whereas initially research on \ac{CP} was driven by the communications community, with a strong emphasis on data transmission efficiency and protocol design, in recent years the field has gained momentum in the robotics and computer vision domains~\cite{wan2025systematic,han2023collaborative,caillot2022survey}. As a consequence, the focus has expanded from merely exchanging raw sensor data or high-level perception outputs to also sharing intermediate neural network features for a wider range of perception tasks. With the advent of 6G technologies, the transmission of large data volumes becomes increasingly feasible, enabling the practical deployment of advanced \ac{CP} methods. This paves the way for more sophisticated, bandwidth-aware cooperation strategies and deeper integration between perception and decision-making processes across cooperative agents.

\subsection{Categories of Collective Perception}

\ac{CP} can be classified into four categories based on the type of data shared between agents: \textit{early}, \textit{intermediate}, \textit{late}, and \textit{hybrid collaboration} \cite{wan2025systematic}.

Early collaboration involves sharing raw sensor data before local processing. This approach preserves the full richness of the original input, enabling high-accuracy perception when data is fused across multiple agents. For example, Ahmed et al. \cite{JointPerceptionScheme-2022-ahmeda} propose a joint perception framework that transmits and fuses LiDAR point clouds prior to detection. While this method offers accurate perception results, it incurs significant communication overhead due to the large data volume.

In contrast, late collaboration transmits only high-level perception results --such as detected objects-- which significantly reduces bandwidth consumption. Xu et al. \cite{ModelAgnosticMultiAgentPerception-2023-xub} present a model-agnostic multi-agent perception system where detection outputs from cooperative agents are transformed into the ego vehicle’s coordinate frame and fused with its own results. This approach, which is what current V2X standards are adopting, is efficient and compatible with heterogeneous sensor setups, but it sacrifices the contextual richness of raw data and is highly sensitive to localization errors among agents.

To strike a balance between data richness and communication cost, intermediate collaboration has been introduced. This method involves sharing intermediate neural network features, which retain more informative representations than final outputs while being more bandwidth-efficient than raw sensor data. A representative example is V2X-ViT by Xu et al. \cite{V2XViTVehicletoEverythingCooperative-2022-xub}, which leverages an attention mechanism \cite{vaswani2017attention} to fuse intermediate features from multiple agents before decoding them into detection results. This approach shows robustness to localization inaccuracies and maintains strong perception performance.

Building on this, hybrid collaboration combines intermediate and late strategies to further optimize bandwidth usage. Liu et al. \cite{RegionBasedHybridCollaborative-2024-liu} introduce a region-based hybrid framework that applies late fusion in non-overlapping fields of view and intermediate fusion where agents share overlapping perception zones. This selective fusion strategy maintains the advantages of intermediate collaboration while reducing the communication load.

In summary, each collaboration type presents a different trade-off between perception performance and bandwidth efficiency. The selected approach should align with system requirements, taking into account factors such as communication constraints, sensor heterogeneity, and localization accuracy.





\subsection{Approaches to Address Collective Perception Challenges in Realistic Scenarios}
While \ac{CP} holds great promise, its real-world deployment is hindered by several practical challenges.

\textit{Localization errors.} They refer to the inaccurate positioning of vehicles in realistic scenarios due to degradation of \ac{GNSS} signal or the lack of \ac{HD} maps covering the driving area. They directly affect the accuracy of \ac{CP} as the accurate spatial alignment is essential for data fusion. To tackle this issue, Song et al. \cite{CooperativePerceptionSystem-2023-songc} proposed the utilization of optimal transport theory to correct the inaccurate vehicle locations and headings. Object-level bounding boxes are utilized as the landmark for localizing the vehicle. Based on the assumption that cooperative vehicles perceive consistent scenarios, transformation errors can be minimized to produce the accurate relative pose between cooperating agents. This object-level approach is communication-efficient, but relies on accurate perception results. With the increase of noise in perception results, its performance degrades.

\textit{Time latency.} Another major obstacle is aligning data from multiple agents to a common time frame, which is essential for fusion. Latency may stem from communication delays, asynchronous processing, or differing sensor update rates. A common mitigation strategy involves predicting object positions at a synchronized time step. For instance, Su et al. \cite{3DMultiObjectTracking-2023-sub} apply a constant acceleration motion model to estimate future positions. Although simple and computationally efficient, this method relies on precise velocity estimates from upstream modules, limiting its robustness.

\textit{Communication interruptions.} 
Packet loss due to network collisions, can severely disrupt \ac{CP} performance. Continuous and reliable data exchange is critical for maintaining perception accuracy. To enhance resilience, Ren et al. \cite{InterruptionAwareCooperativePerception-2024-renc} introduce an interruption-aware framework that predicts missing frames using temporal information from past inputs. However, this approach struggles with long-duration outages and becomes less accurate when historical data is sparse or noisy.


\textit{Heterogeneity in sensors and models.} This aspect further complicates scalable deployment. Features generated by one agent may be incompatible with others due to architectural or modality differences. Xu et al. \cite{BridgingDomainGap-2023-xua} tackle this with the \ac{MPDA} framework, which translates features from a source to a target domain, facilitating inter-agent cooperation. To support multi-modal systems, Lu et al. \cite{ExtensibleFrameworkOpen-2024-luc} propose 
a framework that aligns features from LiDAR and camera modalities in a unified representation space, enabling seamless collaboration despite sensor diversity. Different type of sensors or perception algorithms can also result in \textit{domain shifts}. To tackle domain shifts, Li et al. \cite{DIV2XLearningDomainInvariant-2023-xiangb} introduce a domain-invariant representation through a distillation-based method to bridge gaps between different LiDAR sensor types, enabling more consistent perception across agents. 

\textit{Adversarial attacks.} They pose serious cybersecurity threats in \ac{CP} systems. While connectivity enables cooperation, it also increases vulnerability to malicious inputs. An adversarial message received and used in fusion could compromise vehicle safety. To address this, Li et al. introduce ROBOSAC~\cite{UsAdversariallyRobust-2023-lia}, a framework designed to reach consensus among cooperative agents, mitigating the effects of manipulated data and enhancing resilience against adversarial interference.

In summary, the deployment of \ac{CP} in real-world scenarios faces multifaceted challenges. Nevertheless, ongoing research is actively addressing these issues, pushing \ac{CP} systems toward greater robustness, efficiency, and security. These developments are crucial steps toward realizing the full potential of \ac{CP} in advancing \ac{CCAM}.

\label{tarsce}

\section{Future of Mobility: Key Use Case Groups}
\label{tarsce}

This section outlines the key target scenarios for CCAM systems, summarized in Table \ref{Tab:scenarios} and depicted in Fig.~\ref{fig:CCAM}. The described use case groups cover a range of driving environments from urban intersections to highway support, with an emphasis on the current state-of-the-art and the emergent issues.



\begin{table*}
\caption{CCAM target scenarios: main objectives and relevant challenges.}
\label{Tab:scenarios}
\footnotesize
\centering
\begin{tabular}{p{2.3cm}p{6.8cm}p{7.9cm}}
\hline \hline
\textbf{Scenario} & \textbf{Objectives}& \textbf{Challenges}  \\
\hline
VRU protection &
\begin{itemize}
\item Improve safety for pedestrians, cyclists, and motorcyclists.
\item Extend vehicle perception beyond line-of-sight and sensor range.
\item Enable proactive collision avoidance in complex urban environments.
\end{itemize}
& 
\begin{itemize}
\item Energy efficiency and battery life of active VRU devices (e.g., wearables, smartphones).
\item Scalability and channel congestion with a high density of broadcasting devices.
\item Accurate detection and intent prediction of passive VRUs (non-connected users).
 \item Standardization and interoperability between different types of VRU devices and vehicle systems.
 \item Handling the diverse dynamics of new micro-mobility forms (e.g., e-bikes, e-scooters).
\end{itemize}
\\
Platooning &   
\begin{itemize}
 \item Increase road capacity and improve traffic flow via reduced inter-vehicle gaps.
 \item Reduce fuel consumption and emissions through improved aerodynamics.
 \item Enhance driver comfort and enable new commercial models (e.g., driver rest in follower vehicles).
 \item Enable distributed processing and computational load sharing within the platoon.
\end{itemize}
&

\begin{itemize}
\item Ensuring string stability to prevent shockwave amplification in the platoon.
\item URLLC for vehicle coordination.
\item Commercial viability and demonstrating sufficient fuel savings in real-world conditions.
\item Safe and robust operation in mixed traffic with non-connected vehicles.
\item Transitioning from highway to complex urban environments with ad-hoc platoons.
\end{itemize}
\\ 
Intersection management & 

\begin{itemize}
\item Reduce congestion, delays, and emissions at intersections.
\item Eliminate or minimize stops for connected vehicles, improving energy efficiency.
\item Increase safety by resolving traffic conflicts through coordination.
\item Replace physical infrastructure (traffic lights) with virtual coordination.
\end{itemize}

&

\begin{itemize}
\item Predicting intentions and trajectories of non-connected vehicles reliably.
\item Achieving consensus and fault-tolerant coordination in fully distributed systems.
\item Handling the high complexity and uncertainty of unsignalized intersections and roundabouts.
\item Ensuring safety under all conditions, including communication failures.
\item Integrating with existing infrastructure and traffic management systems.
\end{itemize}
\\ 
Tele-operated driving & 
\begin{itemize}
    \item Enable remote operation of vehicles in complex or constrained environments.
    \item Provide fallback mechanism when autonomous systems fail or encounter edge cases.
    \item Support remote fleet management and last-mile delivery.
    \item Bridge the gap between manual control and full autonomy.
\end{itemize}
& 
\begin{itemize}
    \item Ensuring ultra-low latency and high reliability for real-time control.
    \item Maintaining situational awareness of remote driver and safe operation under network variability.
    \item Designing intuitive and effective \ac{HMI}.
    \item Addressing legal, ethical, and regulatory frameworks for remote driving.

\end{itemize}
\\

Multi-modal CCAM & 
\begin{itemize}
\item Create a seamless, efficient, and integrated mobility ecosystem.
\item Improve safety and coordination between diverse transport modes (cars, trams, drones, robots).
\item Optimize urban logistics and enable new services like coordinated last-mile delivery.
\end{itemize}

&

\begin{itemize}
\item Extreme heterogeneity in vehicle dynamics, communication capabilities, and operational domains (air, ground, rail).
\item Developing common communication protocols and semantic understanding across different domains.
\item Orchestrating complex interactions between autonomous systems from different stakeholders.
\item Ensuring safety and regulatory compliance in shared spaces (e.g., airspace for drones).
\end{itemize}

\\ 

\hline 
 \hline
\end{tabular}
\end{table*}

\subsection{VRU Protection}\label{vru_management}

As urban areas become increasingly automated, protecting VRUs such as pedestrians, cyclists, and motorcyclists becomes a key concern. CCAM systems can enhance safety by enabling vehicles to detect VRUs and exchange safety-related messages via \ac{V2X} communication~\cite{lobo2022enhancing}. 
The VRU detection approaches can be broadly classified based on the level of their integration into the CCAM ecosystem as passive or active \cite{lobo2022enhancing}. 

The \textit{passive VRU} case refers to the individuals who are not equipped with the V2X communication devices (e.g., unprotected pedestrians) and can become visible to the vehicles either via their local sensors or thanks to the \ac{CP}, which is often envisioned to be supported by the road infrastructure aimed at enabling the VRU safety. \textcolor{black}{In \cite{10934058}, for example, Viterbo et al. investigate the use of an infrastructure equipped with sensors to protect \acp{VRU}.}

The \textit{active VRU} case means that the individuals are equipped with devices such as wearables or smartphones. The level of V2X communication support integrated in such devices can greatly vary. In the most restrictive case, the device itself does not support any CCAM-related functions and the vehicles need to process the signals that happen to be transmitted by them in order to detect the VRU on the road and localize it. The interpretation of the situation as well as the scalability of the system in case of many devices in the area are challenging aspects. In more advanced active VRU cases the device explicitly communicates its location and status either directly to the vehicle using one of the dedicated V2X technologies or via the cellular infrastructure where the traditional smartphone can be connected to. Energy efficiency aspect of these approaches requires careful design since persistent broadcasting of the update heartbeat messages might be battery draining. 
Moreover, it can increase the channel load in scenarios with large numbers of VRUs.
To prevent network congestion, VRU clustering has been proposed which can group VRUs in close proximity with similar
kinematic behavior, so that only one of them will broadcast a message as a representative \cite{lobo2023cluster}.

In more futuristic approaches, the VRU communicates not only its status, but also its intentions which could be forecast using a vast amount of the context information.

Finally, recent efforts try to incorporate other types of VRU related to new forms of micro-mobility e.g., e-bikes \cite{campolo2024understanding}. These not only have different dynamics in comparison to pedestrians and bicycles, but also open possibilities for direct management of the engine support based on the communication with the traffic lights and other CCAM infrastructure.


\subsection{Platooning}
\label{platooning}

Historically, platooning has been one of the first use cases envisioned for automated driving systems on highways, where closely spaced trucks communicate using V2X to synchronize their movements~\cite{tsugawa2016review}.
The platoon is defined as a longitudinal formation of vehicles driving together as one unit. While initially envisioned for highway driving, platooning also encompasses urban platooning, where vehicles operate in dense city environments with frequent stops, intersections, and vulnerable road users. It is crucial that the vehicles in the formation are communicating 
in order to efficiently coordinate their movement.
The continuous exchange of information allows vehicles to drive closer than under normal operations 
and this reduces the air resistance;  as a consequence, not only road use efficiency is increased, but also fuel consumption and emissions are reduced. 
Although in recent years platooning has faced some criticisms related to its commercial viability, questioning if the fuel saving benefits in  realistic driving settings can be sufficiently high to motivate the deployment, increased automation level of the formation is still clear and currently prevalent argument for its wide adoption.


From a technical perspective, the automation level of platoon reflects the progress of the cruise control systems. 
In a traditional \ac{CC} a vehicle automatically maintains the speed chosen by the driver. If the CC is complemented by distance measurements, obtained from the onboard sensors, e.g., the frontal radars, then a vehicle is able to maintain the preselected distance or timing to its predecessor. This results in an \ac{ACC} system which is a first step towards automatically maintaining the platooning formation. However, when the number of vehicles increases, the locality of the information sensing becomes the performance bottleneck -- it leads to an amplification of the reaction times within the platoon. Indeed, since the sensing is limited only to an immediate predecessor, the information about the speed adjustments of each individual vehicle propagates along the vehicles with the delay which is proportional to the length of the entire formation. This issue known as \textit{string stability} -- the ability of vehicles in a platoon to maintain consistent speed and spacing -- can be solved by introducing wireless connectivity between vehicles. This brings us to a \ac{CACC} system, where messages from one vehicle can simultaneously reach all the others. Therefore, the V2X-based CCAM platooning systems have an important feature of being able to improve the maneuvering synchronicity of the vehicles in the formation what in the end improves the traffic flow.

Further evolution of the system naturally goes into the direction of increasing the levels of automation for the platooning vehicles. Indeed, enabling not only a longitudinal but also a lateral control would allow reducing the load of the drivers to the level when they are not required to monitor the driving environment all the time. Ultimately, they will even be allowed to sleep, which brings a commercial interest to the platooning back, due to a clear labor cost saving potential. For instance a short formation when the first (leader) vehicle is manually-driven while few others (followers) are autonomous is a practically attractive state-of-the-art use case~\cite{yang2025model}. 


\subsection{Intersection Management}
\label{intersection_management}

Intersection management is critical for urban mobility, where CCAM technologies can significantly enhance traffic flow and safety. 
Current solutions for improving traffic flow at signalized intersections primarily focus on enhancing driver awareness through V2X communication, such as the \ac{GLOSA} \cite{GLOSA}. In this scenario, vehicles communicate with infrastructure to adjust their speed and optimize the timing of traffic lights, reducing congestion and emissions. Advancements are being made toward more adaptive systems that adjust traffic light phases dynamically based on real-time traffic data \cite{adaptive_traffic_lights}. Currently, \ac{C-GLOSA} is under development, which would incorporate the information about real-time queuing at traffic lights to further synchronize the green light timing across multiple vehicles, enabling a smooth, coordinated flow of traffic \cite{cooperative_GLOSA}. 

Another emerging extension of GLOSA is the concept of coordinated (group) start~\cite{etsiTR103578}, where multiple vehicles that are grouped together at an intersection are coordinated by traffic lights. This use case combines platooning and intersection control, where queued vehicles are started synchronously by stabilizing the reaction time in the formation. This approach could significantly reduce congestion
and improve traffic flow, especially in high-density urban environments.

Beyond signalized intersections, cooperative maneuvering also applies to  roundabouts, which present unique challenges due to continuous flow, merging behavior, and limited visibility. Furthermore, disruptive approaches that replace traditional traffic signals with \ac{V2X} communication are also considered under intersection management. As a first approach, \textit{virtual traffic lights} could be used to define which road has priority at an intersection that 
is not equipped with infrastructure \cite{virtual_traffic_lights,JAME2022167}. Further increasing complexity but possibly providing a significant improvement in terms of efficiency, vehicle  could exchange at non-signaled intersections data about their speed, trajectory, and intentions to determine which vehicles should reduce their speed, break, or yield, eventually stopping only when really necessary \cite{10575944,ZHANG2019108563}.
In such scenarios, vehicles communicate in real time and optimize their movements, minimizing stops and reducing traffic congestion. However, challenges remain in ensuring that vehicles can predict and react to the movements of other vehicles, particularly at complex intersections and roundabouts.

Roundabouts pose a unique challenge to automated vehicles due to their dynamic nature and the need for precise vehicle mobility management. Advanced CCAM systems would need to predict the intentions of inbound vehicles, accounting for possible maneuvers such as exiting the roundabout or staying on the main flow. Algorithms would need to incorporate both perception and V2X communication to improve decision-making in real-time. Recent studies (e.g., \cite{roundabout_AI}), demonstrate how AI-driven systems can predict roundabout exits based on vehicle position, heading, and speed. These predictions allow for smoother and safer roundabout navigation for automated vehicles. However, such systems must overcome limitations related to perception errors, V2X communication reliability, and complex traffic environments.

\subsection{Tele-Operated Driving}

\ac{ToD} enables remote (human- or machine-based) control of vehicles, serving as a critical fallback mechanism when autonomous systems encounter limitations in complex or unpredictable environments. This use case is particularly relevant for constrained urban settings, adverse weather conditions, construction zones, and ambiguous traffic scenarios. It also supports remote fleet management and last-mile delivery operations.

Recent experimental evaluations have demonstrated the feasibility of \ac{ToD} over 5G networks. For instance, a study in 
\cite{testouri20255g} used a real-world testbed with real vehicle controlled by a  remote platform to assess key performance metrics, like \ac{RTT} and \ac{G2G} latency. Results show that safe teleoperation can be enforced provided that low vehicle speeds are maintained due to the measured high \ac{G2G} latency. 
Furthermore, the Chair of Automotive Technology at TU Munich \cite{tumteleop} has developed  and explored safety concepts, \ac{HMI}, and legal frameworks for scalable deployment of \ac{ToD}. Their work emphasizes the role of functional safety, visualization strategies, and operator workload management in real-world applications.

A socio-technical analysis by Müller et al. \cite{mueller2024teleop} examines how tele-operation impacts reliability, system complexity, and traffic performance. This research underscores the need for systemic design approaches that account for tele-operator cognition, network variability, and multi-vehicle supervision.

\ac{ToD} thus represents a vital ``bridge'' use case towards  CCAM systems, and requires robust communication infrastructure to ensure safe, scalable, and responsive operation.

\subsection{Multi-Modal CCAM}
\label{multi_modal_ccam}


The integration of multi-modal transport systems, such as automated shuttles, trams, trains, drones, and robots, is an exciting frontier for CCAM. A connected, seamless system would allow various modes of transportation to operate in concert, improving overall efficiency and mobility.

One such example of multi-modal integration is virtual coupling in trains~\cite{wu2023railway}, similar to platooning in trucks. Automated trains could communicate to maintain optimal distances, improving safety and efficiency. Generally, V2X communication between the road and the railway traffic is important, for example, for the improved safety of level crossing.

Drones and autonomous robots are becoming increasingly important in logistics and urban mobility. These systems could integrate with vehicles and infrastructure, sharing data to improve navigation. For instance, drones can be used to improve network connectivity~\cite{chughtai2024drone} as well as to 
provide last-mile delivery solutions in urban environments~\cite{atieh2025impact}. Leveraging V2X communication between the air and the ground to enable \ac{CP} would allow contributing to improving the transportation efficiency and increasing the level of automation. 

CCAM  can enable seamless multi-modal transportation by linking vehicles, infrastructure, and transport services into a unified, intelligent network. Through real-time multi-modal data exchange and predictive coordination~\cite{kueppers2024v2aix}, CCAM can allow different modes of transport (e.g., private cars, automated shuttles, public transit, and freight vehicles) to operate collaboratively rather than independently. This would support dynamic route optimization, synchronized transfers, and efficient resource sharing across modes.

\section {What is Missing: Key Enabling Technologies}
\label{keyenab}


\begin{table*}[t]
\caption{Enabling technologies for CCAM: main benefits, challenges and possible solutions. Part 1: Communication techniques and frequency bands}
\label{Tab:enablers1}
\scriptsize
\centering
\begin{tabular}{p{1cm}p{4.4cm}p{5.2cm}p{5.2cm}}
\hline \hline
\textbf{Enabler} & \textbf{Benefits}& \textbf{Challenges} & \textbf{Solutions} \\
\hline
FD &  \begin{itemize}
    \item Potentially doubled capacity and halved delay
    \item Increased reliability for broadcast transmissions over sidelink thanks to SIC-enabled collision detection while transmitting
\end{itemize} & 
\begin{itemize}
    \item     SI cancellation due to time-dispersive nature and fluctuations of the SI channel under mobility of transmitters, receivers and nearby scatterers 
\end{itemize}
     
    & 
\begin{itemize}
    \item Improved compact antenna design
    \item Powerful transceivers
\end{itemize} \\ 
NOMA & 
\begin{itemize}
  \item Massive connectivity
 \item Improved reliability thanks to successive interference cancellation at the receiver
\end{itemize}
& \begin{itemize}
    \item Imperfect SIC
    \item Processing overhead and delay
\end{itemize}  & \begin{itemize}
    \item Powerful computing capabilities on board the vehicles
    \item Properly tuning the number of cancellation iterations
\end{itemize}  \\ 
RIS & 
\begin{itemize}
    \item Enhanced coverage
    \item Less expensive than deployment of \acp{BS} and \acp{RSU}
    \item Energy efficiency
\end{itemize}
& 
\begin{itemize}
    \item Unintended interference due to imperfect beamforming and specular reflection suppression
    \item Complex channel estimation, in particular for high mobility scenarios
    \item Control and signaling overhead

\end{itemize}
&
\begin{itemize}
    \item Hybrid (active \& passive) RIS deployment
    \item Deploying hierarchical/distributed control architectures 
    \item AI-based beamforming and control
\end{itemize}\\

THz communications & 
\begin{itemize}
    \item Large bandwidth, high achievable data rates
    \item Compact antenna arrays with a large number of elements
    \item Accurate sensing capabilities
\end{itemize}
& 
\begin{itemize}
    \item High propagation losses
    \item High sensitivity to blockage
    \item Rapid channel fluctuations in mobile scenarios
\end{itemize}
&
\begin{itemize}
    \item Beamforming based on non-diffractive beams
    \item Beam alignment and tracking relying on prediction mechanisms and sensing information
    \item Multi-connectivity techniques exploiting multiple links and frequency bands
\end{itemize}
\\ \hline
 \hline
\end{tabular}
\end{table*}

\begin{table*}[t]
\caption{Enabling technologies for CCAM: main benefits, challenges and possible solutions. Part 2: Sensing and cooperative/collective perception}
\label{Tab:enablers2}
\scriptsize
\centering
\begin{tabular}{p{1cm}p{4.4cm}p{5.2cm}p{5.2cm}}
\hline \hline
\textbf{Enabler} & \textbf{Benefits}& \textbf{Challenges} & \textbf{Solutions} \\
\hline

ISAC & \begin{itemize}  
\item 
Reduced channel load and improved network efficiency 
\item Reduced hardware redundancy, leading to lower cost
\item 
Robust in fog, rain, or snow, complementing GPS and camera-based systems
\end{itemize}
& \begin{itemize}
    \item Interference due to shared spectrum
    \item Precise time synchronization requirements  
especially in high-speed environments
\item Standardization across manufacturers and other bodies 
and compatibility with existing networks 
\end{itemize} 
& 

\begin{itemize}
    \item Advanced signal processing and dynamic spectrum allocation techniques to counteract interference
    \item Beamforming and MIMO techniques to 
    separate sensing and communication signals
    \item 
  Network-assisted synchronization 
  to ensure precise timing
\end{itemize} 
\\ 
Localization  &\begin{itemize}
    \item Support full 
    \ac{AD} and platoon/fleet formation 
    \item Enable predictive collision avoidance by early trajectory correction 
    \item Improve efficiency in fleet-wide path planning and dynamic correction 
\end{itemize}  & 
\begin{itemize}
    \item Guaranteeing accurate position suitable for inter-vehicle coordination 
    \item Ensuring seamless positioning in large indoor and mixed indoor-outdoor scenarios 
    \end{itemize} 
& \begin{itemize}
    \item Positioning support enhancements (e.g., within 3GPP) 
    \item Integration of multiple Local / Personal Area wireless technologies 
    with enhanced ranging/positioning capabilities
    \item Improved cooperative positioning through multiple interfaces for more accurate position/distance/angle information
    \end{itemize} 
    \\  \hline 
 \hline
\end{tabular}
\end{table*}

\begin{table*}[t]
\caption{Enabling technologies for CCAM: main benefits, challenges and possible solutions. Part 3: AI \textcolor{black}{and ML} techniques}
\label{Tab:enablers3}
\scriptsize
\centering
\begin{tabular}{p{1cm}p{4.4cm}p{5.2cm}p{5.2cm}}
\hline \hline
\textbf{Enabler} & \textbf{Benefits}& \textbf{Challenges} & \textbf{Solutions} \\\hline    
PQoS &  
\begin{itemize}
    \item 
Autonomous traffic systems enabled to take actions before reaching harmful or critical state
\end{itemize} 
    & 
\begin{itemize}
    \item Wireless links with limited reliability
    \item Predictions for some use cases need to be done seconds, if not minutes, in advance
\end{itemize} & 

\begin{itemize}
    \item Robust ML-based prediction algorithms with safety guarantees
\end{itemize} \\
Distributed AI & \begin{itemize}
   \item Limited raw data exchanges thanks to  exploitation of locally stored datasets
  \item   Easier on-vehicle decision-making; reduced signaling overhead during decision-making %
\end{itemize} & 
\begin{itemize}
    \item Data and device heterogeneity
    \item Dynamic network environments with time-varying conditions affecting communication performance
    \item  AI-related security risks
    \item Poor interpretability of AI models functioning
\end{itemize}   & \begin{itemize}
    \item Data privacy aware techniques (e.g., differential privacy, encryption, autoencoders) to enhance system’s reliability
    \item Adversarial training and anomaly detection safeguard distributed AI models against adversarial threats
    \item explainable AI
\end{itemize} \\
AIaaS & \begin{itemize}
   \item Coordinating discovery of AI-related resources within the CCAM eco-system
   \item Enriching edge/cloud resources with mobile far edge nodes (cars, \acp{VRU}) for CCAM 
\end{itemize} & 
\begin{itemize}
    \item Heterogeneous input data and compute and sensing resources within the CCAM ecosystem 
    \item 
   Ensuring real-time coordination 
 and 
 managing potential network disruptions, also due to mobility and compute overload, or faulty model
updates 
  \item Meeting functional
safety and regulatory constraints
\end{itemize}   & \begin{itemize}
    \item Collective perception and sensor alignment exposure
    
    \item Orchestration mechanisms meeting timing constraints, handling mobility dynamics, e.g, through predicted
trajectories, resource availability, and radio conditions

    \item Model metadata exposure, certification tracking, and runtime
monitoring 
\end{itemize} \\
\hline
 \hline
\end{tabular}
\end{table*}

\begin{table*}[t]
\caption{Enabling technologies for CCAM: main benefits, challenges and possible solutions. Part 4: Virtualization} 
\label{Tab:enablers4}
\scriptsize
\centering
\begin{tabular}{p{1cm}p{4.4cm}p{5.2cm}p{5.2cm}}
\hline \hline
\textbf{Enabler} & \textbf{Benefits}& \textbf{Challenges} & \textbf{Solutions} \\ \hline
 
DTs 
& \begin{itemize}
\item Traffic optimization
\item Hazard detection and prevention 
\item Predictive maintenance
\item Enhanced safety and risk management
\item System-level decision support
\end{itemize} & \begin{itemize}
\item Sensing and modeling accuracy
\item Model interoperability and standardization
\item Low-latency, high-reliability physical-digital communication
\item Computation complexity
\item Privacy and security
\end{itemize} & 
\begin{itemize}
\item ML to synthesize multi-source sensor data and to improve model accuracy and resilience 
\item Edge computing to reduce latency and support real-time synchronization 
\item Hierarchical computing to scale DTs  
\item Interoperable frameworks and standards (APIs, data formats, protocols) to facilitate integration across vendors and systems
\item Embedding encryption, access control, and anomaly detection 
to ensure data integrity and compliance 
\end{itemize} \\
SDIoV & 
\begin{itemize}
    \item 
Centralized, programmable network control and
\item Low-latency decision-making through integration with edge and fog computing 
\end{itemize}

& 
\begin{itemize}
    \item Managing heterogeneous networks
\item Maintaining control under high mobility
\item Security issues
\end{itemize}

& 

\begin{itemize}
    \item Dynamic SDN-based control
    \item Localized processing via edge and fog nodes
    \item Collaborative resource sharing among vehicles and infrastructure
\end{itemize}
    \\
\hline 
 \hline
\end{tabular}
\end{table*}

\begin{table*}[h!]
\centering
\caption{Key Enabling Technologies and Their Relevance for Scenarios identified in Section~\ref{tarsce}} 
\begin{tabular}{p{3.5cm}p{6cm}p{6cm}}
\hline \hline
\textbf{Technology} & \textbf{Main Scenarios} & \textbf{Remarks}\\
\hline
FD & Platooning (including urban), Intersection Management & Improves reliability of broadcast messages in dense environments\\\\
NOMA & VRU protection, Platooning (including urban) & Enhances connectivity in congested networks \\\\
RIS & Intersection Management (including roundabouts), Urban Platooning & Helps mitigate LOS blockage in urban settings \\\\
THz Communications & Multimodal CCAM, VRU Detection & Enables high-throughput data exchange for sensor-rich environments\\\\
\ac{ISAC} & All scenarios, especially VRU Detection, Intersection Management, Roundabouts & Especially relevant for VRU detection, intersection management, and cooperative maneuvers\\\\

 Localization & Platooning (including urban), Intersection Management & Supports precise positioning and coordination\\\\
PQoS  & Platooning, Tele-operated Driving & Ensures reliable communication for coordinated maneuvers \\\\
Distributed AI & All scenarios, including Tele-operated Driving & Supports real-time decision-making and privacy-preserving data sharing \\\\
AIaaS & All scenarios & Provide assistance through exposure APIs for complex AI orchestration within CCAM entities \\\\
DTs & Multimodal CCAM, Platooning (including urban), Tele-operated Driving & Simulates and optimizes traffic and vehicle behavior\\\\
SDIoV & All scenarios & Provides flexible, programmable infrastructure for CCAM coordination \\\\
\hline \hline
\end{tabular}
\label{tab:tech_scenarios}
\end{table*}

Despite the recent efforts in standardization bodies to improve the performance of V2X communication technologies, the challenging demands of the aforementioned \ac{CCAM} use cases require investigation of further innovative, even disruptive, technological enablers to be identified also in the 6G realm. 

In the following, we first review the main ongoing efforts in defining 6G. 
Then, the main enablers for \ac{CCAM} are discussed, 
along with the main challenges to be addressed 
and possible solutions. A short summary of the discussed technologies is available in  
Table~\ref{Tab:enablers1} to Table~\ref{Tab:enablers4}, 
whereas the relevance of each of the technologies to the scenarios identified in Section~\ref{tarsce} is available in Table~\ref{tab:tech_scenarios}.

\subsection{6G Usage Scenarios and Enablers}

Although it is still unclear how 6G will be shaped, the 
 main players agree that the next generation networks will go beyond their traditional role of  connectivity enablers\cite{fiore2023full},
\cite{tsekenis2024towards}, \cite{kerboeuf2024design}, \cite{liu2025itu}. Indeed, a plethora of novel services and capabilities powered by technologies beyond those supporting mere connectivity will be offered. 

\begin{figure}[t]
\centering
\includegraphics[width=0.85\columnwidth,draft=false]{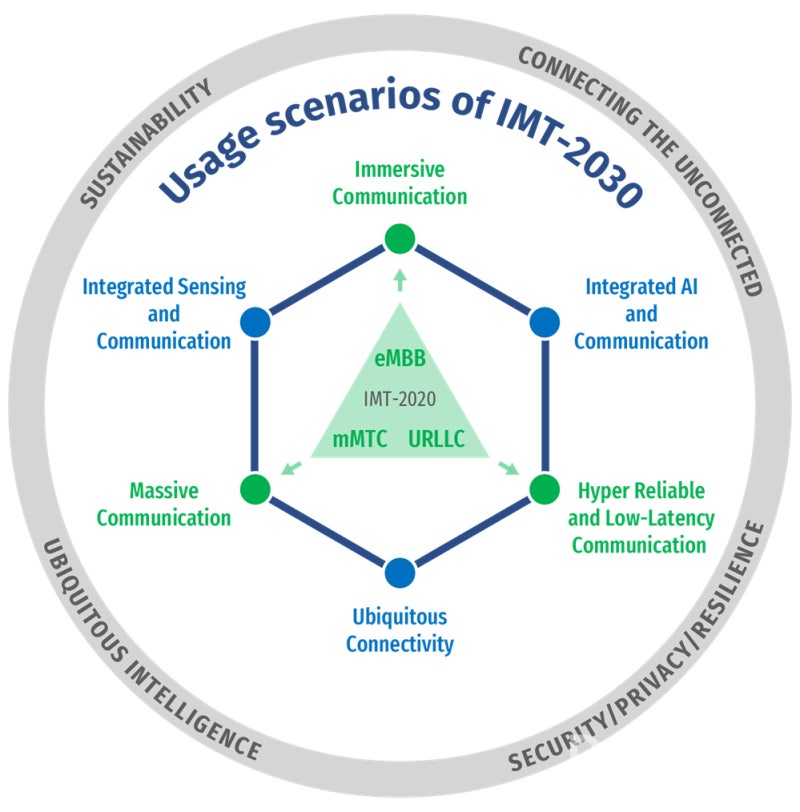}
\caption{The 6G wheel as approved by ITU-R (figure from \cite{itu}).}
\label{fig:wheel}
\end{figure}

That said, 6G is expected to improve traditional communication services as well as those foreseen in \ac{5G}, i.e., \ac{eMBB}, \ac{mMTC}, \ac{URLLC}. Extended usage scenarios, as identified by the \ac{ITU-R}  \cite{itu}, will include immersive communication, massive communication, and \ac{HRLLC}. They respectively cover novel use cases, including \ac{XR}
and holographic communications, remote
surgeries and fully autonomous driving, dense \ac{IoT} deployments in small areas
for industries, metaverse, logistics and transportation \cite{liu2025itu}. 
In addition, there is a group of novel usage scenarios (Fig.~\ref{fig:wheel}) namely,
ubiquitous connectivity, 
\ac{ISAC}, \ac{AI} and communication, that demand unprecedented and also interdisciplinary efforts.
Ubiquitous connectivity is aimed to bridge the digital divide, by offering connectivity particularly in rural, remote
and sparsely populated areas. 
ISAC will encompass new applications benefiting from wide-area, multi-dimensional
sensing that detects the presence and/or movement of both
unconnected objects and connected devices.
6G will also offer services natively integrating AI and communication functions. Moreover, in the so-called 6G wheel (Fig. \ref{fig:wheel}), four overarching aspects
acting as design principles commonly applicable to all usage scenarios are identified which
 need to be considered
when designing 6G systems: sustainability, connecting the unconnected, security/resilience, ubiquitous intelligence.


Several works, e.g., \cite{liu2025itu}, \cite{jiang2021road},
\cite{trichias20246g}, have identified the main 6G technical enablers. Among them, some 
 span the lower protocol layers and mainly refer to the communication domain,  e.g., \ac{FD}, \ac{NOMA}, \ac{RIS}.  Others, instead, are more transversal such as \ac{ISAC} techniques, \ac{AI} algorithms, \ac{DT}. 
Those deemed more relevant in the CCAM context are analyzed in the following.




\subsection{Communication Techniques and Frequency Bands}


 As the CCAM technology achieves high penetration rates, the available radio spectrum is expected to become increasingly congested. Moreover, the harsh propagation conditions faced under high mobility settings can hinder reliability and timeliness of CCAM services. To address these issues, novel techniques at the \ac{PHY} and \ac{MAC} layers, as well as the exploitation of higher frequency bands, can be employed, as discussed in the following.

\subsubsection{Full-Duplex} Among the key technologies for 6G, a promising one is in-band \ac{FD}, i.e., the ability to transmit and receive simultaneously in the same band, with inherent benefits in terms of larger capacity, higher throughput, lower latency \cite{smida2023full}.
Several studies have considered the application of FD techniques in vehicular environments with focus on the capability of sensing the medium while transmitting. This proved to be greatly beneficial for (unacknowledged) broadcast traffic, e.g., \acp{CAM}
, for which packet losses may detrimentally affect the safety of vehicles. Efforts have been devoted both to overhaul IEEE 802.11p \cite{bazzi2018enhancing} and, more recently, C-V2X distributed access modes
\cite{todisco2024full}, \cite{bankov2024enhancing}, \cite{aslani2024performance}, \cite{ahmad2024coverage}. 

The main idea for IEEE 802.11p is to let an FD
vehicle detect a CAM collision while transmitting, promptly abort
the packet and retransmit it later \cite{bazzi2018enhancing}, to increase the reliability. Similarly, in C-V2X autonomous mode, vehicles detect collisions while transmitting and react to detected collisions through retransmission procedures. In addition, resource reselection is enforced to avoid persistent collisions in the case of periodic traffic, for which the same frequency resources are reserved for several transmission attempts \cite{todisco2024full}.

\textit{Challenges of FD for V2X.} Although FD is more than a decade old, the lack of effective \ac{SFIC} techniques has so far limited its adoption to small-form factor
devices (e.g., consumer handheld devices, like smartphones
and tablets). Indeed, additional hardware/software components necessary for
effective \ac{SFIC} complicate the transceiver
design and consume power and resources~\cite{mohammadi2023comprehensive}. Moreover, issues related to the time-dispersive nature and the fluctuations of the
self-interference channel, caused by the moving scatterers and
reflectors in the surrounding environment and the
mobility of the device itself, are exacerbated in vehicular scenarios. 

\textit{Addressing Challenges of FD for V2X.}
In vehicles, the energy and computing constraints related to the complexity of FD transceivers in handling \ac{SFIC} are less severe than those in small-form-factor consumer devices.
However, there is still room for designing low-complexity algorithms ensuring more robust \ac{SFIC}. 
Antenna design can also play a role to enable the widespread usage of FD technology \cite{ahirwar2024compact}.

\subsubsection{NOMA} Sidelink V2X communications typically rely on techniques aimed at avoiding interference among simultaneously
transmitting vehicles by selecting
non-overlapping resources.
However,
the limitations of channel sensing to assess resource availability and signaling procedures, such as their inability to fully capture the dynamics
of the vehicular environment, combined with the presence of
hidden terminals, prevent from completely avoiding packet collisions.

\ac{NOMA} techniques allow multiple users to utilize overlapping time and frequency
resources concurrently where packets from multiple transmitters can be successfully decoded thanks to \ac{SSIC} techniques.
In particular, in \cite{bazzi2023toward} the authors investigated the use
of advanced receivers that employ SSIC technology to mitigate the interference caused by overlapping transmissions over sidelink when the autonomous mode is used.
This interference becomes particularly
critical under high data traffic conditions, with multiple transmissions (and retransmissions), as expected in scenarios with a
high penetration rate of CCAM applications.
In such scenarios, SSIC techniques are coupled with the legacy autonomous mode of C-V2X. Results show that resolving the collisions by increasing the complexity of the receiver can be more beneficial than
designing sophisticated resource allocation protocols that attempt to avoid
them \textit{a priori}. 

In \cite{rajalakshmi2024enhancing}, power-domain NOMA is used to superimpose
\acp{CAM} and \acp{DENM} on the same sub-channel. A larger power level is assigned to \acp{DENM} compared to \acp{CAM}, given their higher priority, in order to 
enhance the utilization of radio resources.

\textit{Challenges of NOMA for V2X.}
So far, few results have been provided, and under favourable assumptions, such as
ideal SSIC capability and an infinite number of iterations \cite{bazzi2023toward}.
Imperfect SSIC should be considered instead, to properly assess to which extent NOMA can be beneficial. 
Together with quantifying the SSIC error, another practical challenge is the potentially extended SSIC delay and processing overhead incurred at the receiver.
The computational complexity of SSIC requires the definition and/or identification of specific  optimized hardware and software implementations. 

\textit{Addressing Challenges of NOMA for V2X.}
Compared to other domains, NOMA in V2X appears more feasible considering the power availability of vehicles, and given the capabilities of modern on-board computing platforms. The main parameter affecting the processing burden  \cite{almohamad2020low} and the decoding time \cite{manglayev2017noma}, \cite{saito2013non} is the number of iterations; however, results in \cite{todisco2023sic} show that a single iteration already brings substantial performance gains in vehicular scenarios. 

\subsubsection {RIS for V2X} 


The use of \ac{RIS} is envisioned as a key enabling technology in future 6G networks, offering a paradigm shift in how wireless propagation environments can be dynamically shaped to improve communication performance. In vehicular networks, where high mobility, dynamic topology, and frequent \ac{LOS} obstructions challenge the reliability and latency of \ac{V2X} links, \ac{RIS}  can play a fundamental role. By leveraging metasurface elements capable of reconfiguring the phase, amplitude, and polarization of incident electromagnetic waves, in fact, 
\ac{RIS} can enhance signal coverage, mitigate blockages, and facilitate reliable connectivity in \ac{NLOS} conditions. While traditionally envisioned as being deployed on fixed infrastructure such as roadside units, building facades, and traffic signs, \ac{RIS}  can also be integrated onto mobile platforms, including on roofs and doors of a vehicle \cite{10926852}, thus enabling cooperative reflection and beamforming in highly dynamic vehicular scenarios~\cite{naaz2024empowering}.

\textit{Challenges of RIS for V2X.}
The application of \ac{RIS} in vehicular networks presents several challenges. In fact, the fast-changing geometry and Doppler effects due to high mobility complicate real-time beam alignment and phase configuration, requiring ultra fast and predictive control algorithms imposing strict latency requirements on \ac{RIS} configuration and beam alignment \cite{9110912}. In addition, the integration of \ac{RIS} in moving vehicles introduces additional synchronization and coordination complexity, particularly in multi hop and cooperative reflection scenarios. 
While fixed \ac{RIS} can be optimized for strategic locations, mobile \ac{RIS} mounted on vehicles or drones require continuous synchronization and cooperative beamforming strategies to maintain effective reflection links \cite{9148961}. Mobile \ac{RIS} platforms also demand efficient signaling and coordination mechanisms to share environmental and positioning data without overloading the network control plane. Moreover, the Doppler spread induced by moving scatterers and reflectors can degrade the coherence of \ac{RIS}-assisted links.
Finally, channel estimation remains an open issue due to the large number of passive elements and rapidly varying environments. 

\textit{Addressing Challenges of RIS for V2X.}
Traditional methods for real-time channel estimation are computationally expensive and impractical given the passive nature of \ac{RIS} elements and the  number of reflective units. To mitigate this, recent works propose machine learning-based solutions, such as deep reinforcement learning and graph neural networks, to enable fast and adaptive \ac{RIS} control with partial or delayed \ac{CSI} \cite{8741198}. 
To face the issues introduced by moving scatterers and reflectors, recent solutions also include hybrid \ac{RIS} architectures combining passive reflection with a small number of active sensing 
to enable localized environmental awareness and Doppler compensation \cite{9206044}. Additionally, \ac{RIS}-assisted vehicular networks need scalable control architectures, possibly incorporating edge intelligence and \ac{SDN}, to dynamically adapt reflection profiles based on predictive models of vehicle trajectories and environmental maps \cite{9326394}.




\subsubsection{Terahertz Frequency Bands} 
Thanks to the vast available spectrum spanning from 0.1 to 10~THz, operating in the so-called THz frequency bands offers the opportunity to leverage large transmission bandwidth and dramatically boost  communication performance, achieving data rates in the order of terabits per second~\cite{THZBook}. This makes THz communications a good candidate for supporting CCAM applications requiring the transmission of large amount of real-time data, such as sensor information, 3D maps, video and LiDAR streams. At THz frequencies, the short wavelength also allows to realize compact antenna arrays with a large number of elements able to focus the radiated energy into narrow beams. In turn, this helps in reducing the interference and increasing the spatial reuse of the wireless channel, which is particularly beneficial in scenarios with a high density of connected vehicles. Additionally, THz bands offer unique sensing capabilities which makes them suitable for realizing the 
aforementioned \ac{ISAC} paradigm. The large available bandwidth and enhanced directionality enables in fact to achieve highly accurate resolution in the range and angular domains, which is crucial for use cases demanding for precise detection, localization, and imaging of objects. The unique way in which THz waves interact with various substances makes them suitable also for characterizing the dielectric properties of objects and materials and measuring the concentration of gases and pollutants in the air. 

\textit{Challenges of THz bands for V2X.}
The applicability of THz communications in vehicular scenarios is hindered by important challenges. Due to the high propagation losses at such high frequencies, THz links are generally suitable for short-range communications, such as between neighboring vehicles, roadside infrastructure, or within the vehicle itself. Atmospheric effects like rain and snow introduce additional attenuation which may further degrade the performance. Moreover, THz signals are highly susceptible to blockage from neighboring vehicles and obstacles, leading to intermittent connectivity. In high-mobility scenarios, THz links also experience rapid channel fluctuation, requiring fast link adaptation mechanisms.

\textit{Addressing Challenges of THz bands for V2X}.
To address the aforementioned challenges, new antenna and beamforming solutions capable of creating and handling highly directional links are needed. Approaches based on non-diffracting beams, such as Bessel and Airy beams, can help mitigate the impact of blockers~\cite{reddy2023ultrabroadband, lee2025experimental}. Furthermore, the small beam size and fast-changing nature of THz channels necessitate intelligent beam alignment and tracking protocols relying on prediction mechanisms and sensing information~\cite{chen2022enhancing, 10149134}. Finally, novel multi-connectivity techniques that can dynamically exploit multiple links and frequency bands simultaneously offer promising solutions to enhance the robustness of THz communications under harsh propagation conditions \cite{sopin2022user}.



\subsection {Sensing and Cooperative/Collective Perception} 


The convergence of communication, sensing, and localization is emerging as a key paradigm of future 6G vehicular networks. 
\ac{ISAC} plays a central role in this transformation by leveraging advances such as massive \ac{MIMO}, wideband waveforms, and operation at mmWave and sub-THz frequencies to extract sensing information, 
enabling spectrum reuse and more efficient resource management. Complementing ISAC, high-accuracy localization is essential to support advanced CCAM functionalities. This is especially true in GNSS-denied environments such as urban canyons or tunnels, or in applications like platooning and coordinated maneuvers, where sub-meter accuracy and high availability are critical. 


\subsubsection{ISAC}
%


Both monostatic and multistatic 
sensing configurations are possible to enable ISAC with V2X communication. 
Monostatic sensing refers to when the transmitter and the sensing receiver are co-located, 
requiring 
the use of full-duplex radios to enable simultaneous transmission and reception. 
On the other hand, when the transmitter and one or more sensing receivers are not co-located, we have a multistatic configuration. 

In  
the radar community, 
there are no unified standardization rules \cite{9924643}, thus vendors provide different proprietary solutions leading to uncontrolled interference, severely affecting a radar's sensing capability \cite{10734974}. 
In contrast, for transmission of signals in a standardized communication system, resource allocation policies can be employed to control the interference, representing a remarkable advantage for future sensing compared to current monostatic radar sensors.

To achieve a high degree of integration between sensing and communication, V2X networks should  manage the sensing process, schedule wireless resources and process data granting privacy and security, which are not trivial operations when trying to meet communication and sensing requirements. 
Such integration could efficiently use the scarce available resources by sharing spectrum, hardware, and signal processing blocks, by
exploiting signals 
used for V2X communications. In fact, even if the integration can be done at different levels, it will be the sharing of sites, spectrum, hardware, waveforms and the reuse of communication signals for sensing that could allow interesting new applications.

\textit{Challenges of ISAC for V2X.} The most challenging ISAC configuration enabled by V2X is sidelink.  
In particular, it is difficult to achieve accurate device-based positioning with poor network coverage or high congestion and allowing device-free object detection using sidelink signals \cite{BarDecMasGioZanBazSti:J24,DecGueGioGuiMas:J23,PucPaoGio:J22}. V2X communication has been designed to operate in frequency ranges FR1 and FR2, with
most implementations limited to FR1 with bandwidth up to 20 MHz, despite FR1's support for up to 100 MHz. This limitation significantly reduces the available physical resources, thereby increasing the probability of collisions, especially when autonomous resource allocation schemes are utilized.
In FR2, beam management is required for communication 
between vehicle \acp{UE} \cite{Montero:BeamMan5Gsidelink,Fazliu:BeamManUrban}, 
and as it is dependent on the vehicles' position, it provides a strong synergy between sidelink communication and localization. Moreover, ISAC systems may 
exchange sensitive data, making them vulnerable to cyberattacks and privacy breaches that must be kept under control.

\begin{figure}[h]
\centering
\includegraphics[width=0.95\columnwidth,draft=false]{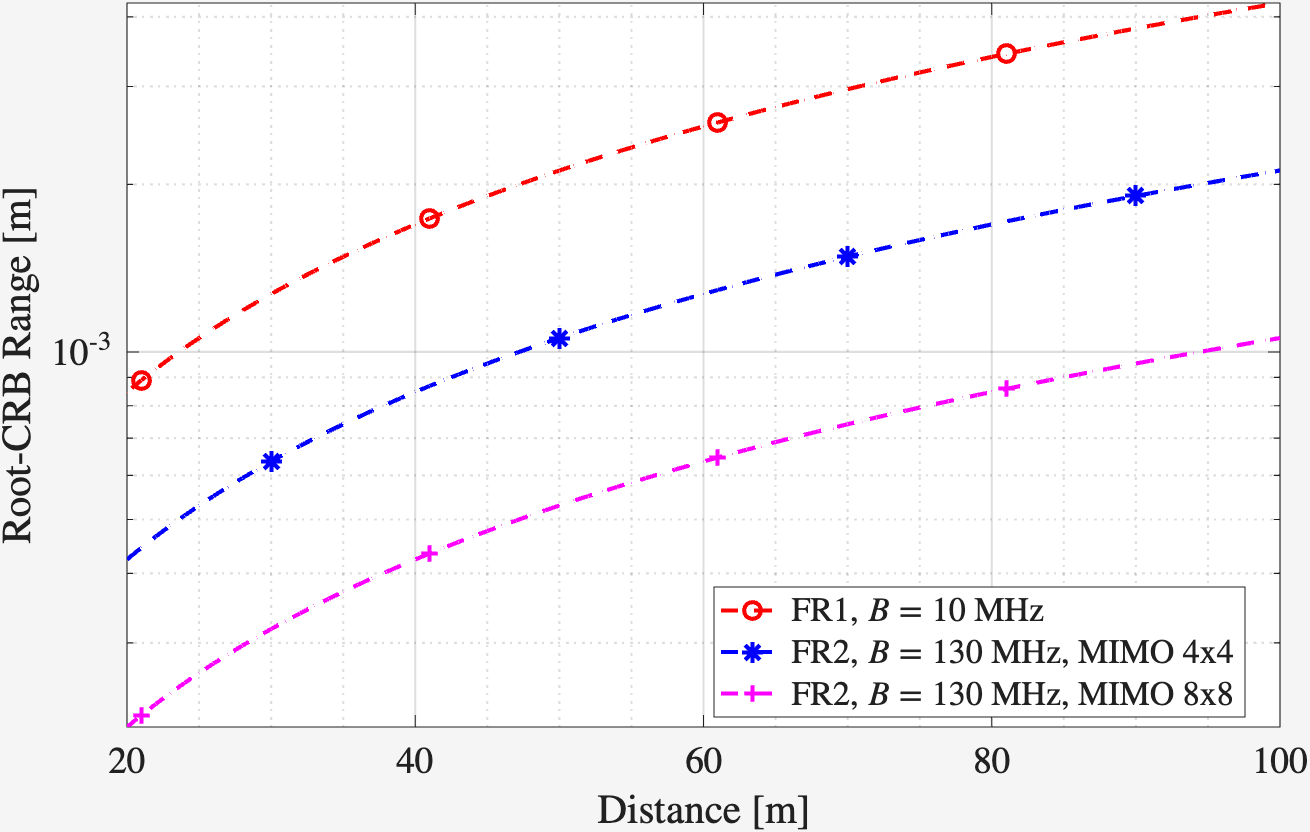}
\caption{Example of CRLB for localization of a passive target. 
}
\label{fig:ISAC}
\end{figure}

\textit{Addressing Challenges of ISAC for V2X.}
%
Target detection, ranging, and positioning, are some of the key elements for ISAC that can greatly benefit from the use of the larger bandwidth (up to 400 MHz) that can be supported in FR2 for sidelink communications.
Hence, the frequency ranges FR1 and FR2 could cover different
use cases. 
In addition, the use of advanced resource allocation schemes could be considered to dynamically assign resources based on real-time demand for sensing and communication. 
For example, in Fig. \ref{fig:ISAC}, the root \ac{CRLB} with monostatic sensing for the range estimation of a target with $1$ m$^2$ radar cross-section (e.g., a vulnerable road user)
is plotted considering different bandwidth and antenna arrays at FR1 and FR2.
Although increasing the bandwidth enhances range estimation accuracy, moving to higher frequency bands 
results in greater path loss, for which 
a \ac{MIMO} setup is essential.

Regarding privacy and security, encryption and authentication mechanisms to protect data integrity and confidentiality are under study using ML-based intrusion detection systems to identify and mitigate potential cyber threats and adopting privacy-preserving techniques like federated learning to minimize the sharing of raw data.

\subsubsection{Localization} 
Localization\footnote{Also referred to as \emph{positioning} in the following.} is by definition a key functionality in the deployment of CCAM solutions: the availability of information on the position of each vehicle is paramount to enable autonomous operations, and the information about mutual position across groups of vehicles enables the application of short-term trajectory planning to predict and prevent collisions before last-second collision-avoidance mechanisms kick in; the availability of position information fleet-wide also enables optimizations in long-term trajectory planning and load balancing for service requests that impact on mobility. Localization is a functionality that has acquired an increased importance across the landscape of wireless technologies, with all major standardization bodies now regularly including positioning features as part of their standardization process. As already mentioned, this is the case for the 3GPP, with 5G-specific positioning features introduced in Rel-16 and further expanded in Rel-17 and Rel-18; but the same trend can be observed in IEEE. In the family of local area network technologies, following on the steps of the 802.11-2016, where Fine Time Measurements were introduced that allow for 1-2 m positioning accuracy, the recent 802.11az standard promises sub-meter accuracy; some of these procedures are already included in the \ac{NGV} mode of IEEE 802.11bd now part of IEEE 802.11-2024 \cite{Car2Car_11bd}. At the same time, personal area network technologies evolve as well toward accurate distance, \ac{AoA} and position estimation, see for example the AoA feature introduced in Bluetooth 5.1  and the planned update for UWB with the  802.15.4ab standard (Next Generation UWB) that integrates communications, positioning and sensing. Position information will thus be no longer provided only by dedicated technologies, such as GNSS, but often in a complementary manner by multiple wireless technologies. This evolution is expected to play a key role in the deployment of CCAM solutions, helping to address the challenges that characterize the acquisition of position information in modern and future mobility-related applications.

\textit{Challenges of Localization  for V2X.}
The introduction of automated mobility will require position information to be seamlessly available to the vehicle to enable automated driving, something that cannot be guaranteed by a single technology: urban canyon propagation conditions or mobility in large indoor areas are examples of scenarios potentially leading to failures or out of service events in GNSS-based positioning. In addition, while the typical positioning accuracy provided by GNSS, in the order of 2-5 meters, is enough to support car navigators thanks to the additional information extracted by maps, this will not be the case for many CCAM scenarios, particularly those requiring accurate information on mutual distances and positions on vehicles moving in formation or platooning. 
Proposals to address this issue in the past typically relied on cooperation between vehicles, based on the exchange and fusion of measurements, but the resulting accuracy was affected by the quality of the measurements made available by the underlying wireless technologies.

\textit{Addressing Challenges of Localization  for V2X.} The expected evolution of positioning features in radio mobile systems discussed above may offer a viable solution to address the continuity and accuracy issues posed by GNSS in CCAM scenarios. In particular, within 3GPP 5G and beyond 5G systems may indeed help in this regard in multiple ways: 
\begin{itemize}
\item Assistance to GNSS positioning: support to GNSS positioning has been part of the 3GPP landscape since Rel-15, and subsequent releases added additional features, such as the distribution of GNSS correction data in broadcast fashion to ensure scalability with increasing user spatial density and provisions to ensure the possibility for the devices to verify the integrity of GNSS data;
\item Radio access technology (RAT)\acused{RAT}-dependent positioning: positioning using the 5G-NR radio interface was introduced in Rel-16, with the introduction of the \ac{LMF} and the definition of positioning algorithms 
capable of leveraging the larger bandwidth of 5G signals. Standardization of accurate positioning was further carried out in subsequent releases, with target performance set to sub-meter accuracy \cite{3gpp22872}.

\item \ac{RAT}-independent positioning: as 3GPP releases progress, support of \ac{RAT}-independent positioning solutions is expected to increase as well, improving the support for positioning based on \ac{WLAN}/\ac{WPAN} technologies like WiFi or Bluetooth and for \ac{RTK}; as the positioning accuracy of such alternative technologies improves, thanks to the trend previously discussed, the role of \ac{RAT}-independent positioning in supporting CCAM applications will become more and more important. 

\item Improved cooperative positioning: the availability of multiple technologies that can provide accurate device-to device information, like the 3GPP sidelink positioning standardized as part of Rel-18, as mentioned in Section \ref{standard}, can play a key role: it will in fact enable autonomous, \ac{UE}-side positioning, bypassing the coverage and latency issues that may affect network-side positioning. Proposals that originally required the availability of vision or LiDAR-based technologies to collect data to be shared across vehicles in a fleet/platoon will take advantage of radio signals providing a similar accuracy \cite{8307478}.
\end{itemize}

\subsection {AI \textcolor{black}{and ML} Techniques} 


\textcolor{black}{\ac{AI} and \ac{ML} are widely recognized as key enablers of next-generation communication networks and future \ac{CCAM} systems. Their role goes beyond improving individual algorithms or isolated network functions: in 6G-enabled \ac{CCAM}, \ac{AI}/\ac{ML} is expected to support the continuous adaptation of communication, sensing, computing, and service-management functions in highly dynamic vehicular environments~\cite{casetti2024ai}.}


\textcolor{black}{In the \ac{6G} vision, intelligence is intended to be ubiquitous as it is needed in every part of the network, from servers and \acp{BS} to end-devices and
sensors. Indeed, it can benefit network operations,
wireless propagation environments, and communication services, leading respectively to operational intelligence, environmental
intelligence, and service intelligence \cite{dang2020should}.}

The ability to process and analyze large amounts of data in real-time coming from on-board and road-side sensors\textcolor{black}{, and also encompassing environmental information (e.g., weather/road conditions, buildings,
traffic density,  moving
obstacles)} enables the involved entities to make smarter decisions, predict and adapt to dynamic road and connectivity conditions in real-time.


\subsubsection{Predictive Quality of Service (PQoS)\acused{PQoS}}

By predicting the variation of the network \ac{QoS} before it occurs, \ac{PQoS}~\cite{boban2021predictive} can play a vital role in the CCAM context. 
Such predictive capability helps maintain high  reliability required for cooperative maneuvers. For instance, if a platoon of vehicles is about to enter an area where network coverage is weak, the PQoS system can predict this drop and take proactive measures. These might include preemptively adjusting the vehicles' following distances, changing the platoon's route, or switching to alternative communication channels to avoid disruption. ML is the \textit{de facto} tool for enabling PQoS in connected cooperative driving~\cite{9771590}. By analyzing existing communication patterns, environmental factors, and the current state of the network, ML can predict future QoS. In cooperative driving, this might involve predicting how network latency will affect the timing of a coordinated lane change or how packet loss could impact the synchronization of braking systems in a vehicle platoon. By predicting these conditions, the system can adjust the communication parameters or alter driving strategies to maintain the seamless coordination of vehicles. This ensures that even in the case of communication challenges, the cooperative driving system remains robust, and vehicles can continue to operate safely and efficiently.

\textit{Challenges of PQoS for V2X.}
While PQoS offers significant advantages for connected cooperative driving, it also introduces new challenges. These include: \textit{(i)} Highly dynamic network conditions with high vehicular mobility and changing surrounding environment; \textit{(ii)} Diverse communication requirements across V2X use cases, including stringent latency and reliability needs and heterogeneous traffic types; \textit{(iii)} Scalability and coordination of PQoS solutions to dense networks with thousands of vehicles; and \textit{(iv)} Processing large amounts of data in real-time, which requires considerable computational power and sophisticated algorithms.

\textit{Addressing Challenges of PQoS fo V2X.} Despite these challenges, the integration of PQoS in connected cooperative driving promises to enhance the safety, efficiency, and reliability of autonomous vehicle operations. 
Using multi-\ac{RAT} connectivity and predictive handover strategies based on trajectory forecasting and channel state prediction can address the highly dynamic nature of vehicular networks. In terms of diverse QoS requirements, employing dynamic QoS-aware traffic prioritization can help meet varying service needs. Leveraging ML models (including deep learning and context-aware prediction) trained on diverse vehicular datasets can address the complex prediction task. 

\begin{figure}[t]
\centering
\includegraphics[width=0.95\columnwidth]{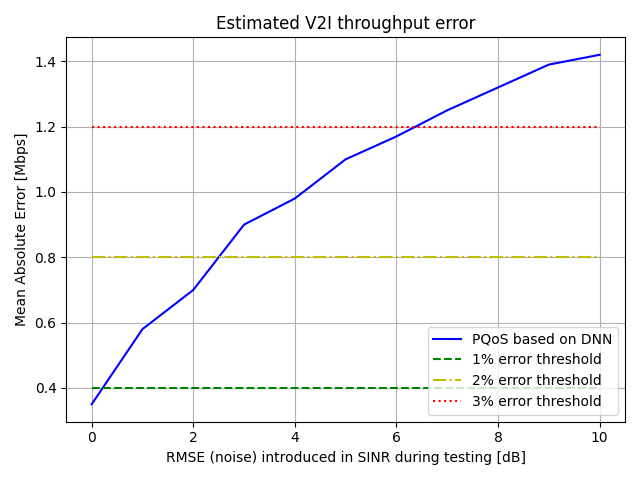}
\caption{Benefit of PQoS in estimating V2I throughput.}
\label{fig:PQoS}
\end{figure}

The use of advanced ML techniques for PQoS shows promise in addressing several of the above challenges. Specifically, potential benefits of PQoS for CCAM related use cases 
have been discussed in~\cite{boban2021predictive}. Fig.~\ref{fig:PQoS} shows the performance gain of a PQoS solution based on Deep Neural Networks (analysis extended from~\cite{boban2022measurement}) in terms of predicting the \ac{V2I} throughput by using \ac{SINR} as the input signal for prediction. 
When the input signal has been measured without additional noise, the \ac{MAE} of PQoS is 0.38 Mbps, which, given that the prediction is on the \hbox{0-40 Mbps} interval, is below 1\% error. Increasing the noise in the input training signal to below 3~dB results in below 2\% error, whereas increasing it above 6~dB results in above 3\% error. These results indicate that, if the input signal is properly measured, PQoS can provide highly accurate throughput prediction.

Overall, by shifting from a reactive to a proactive approach, PQoS ensures that connected vehicles can anticipate and adapt to changing network conditions, thereby maintaining the continuous and reliable communication necessary for complex cooperative maneuvers. Therefore, PQoS can become a crucial element in the success of connected cooperative driving, as it enables vehicles to maintain the high levels of communication required for safe and efficient coordination.

\subsubsection{Distributed AI} 
Traditionally, centralized AI approaches have been employed to V2X communications, requiring data transfers from source devices to centrally located data centers and, thus, suffering from data privacy, computational complexity and operational costs \cite{one6G_Whitepaper_overview}. Alternative AI solutions have also recently emerged (e.g., Split Learning, Federated Learning), distributing AI related tasks of reduced computational complexity (e.g., local training) to a number of (in general heterogeneous) devices, while exploiting locally stored datasets and avoiding any raw data exchanges \cite{NASSEF2022108820}.

In V2X environments, distributed AI approaches can be of value in a number of ways \cite{electronics12194023, Banitalebi9908149, Liu7922594, Magoula9977782}. Firstly, by transferring the AI logic closer to the data generation, data privacy can be preserved. \textcolor{black}{This is crucial in cases where highly sensitive or private information, such as vehicle location, driving behavior, or personal data, may pose restrictions in data exchange.} Secondly, distributed solutions could significantly reduce signaling overhead by performing computations locally at the edge, decreasing the amount of data exchanges, while optimizing the use of available network resources. This benefit could be especially relevant in high density urban traffic scenarios, where the communication channel occupancy increases proportionally to the number of computing devices.  Furthermore, distributed AI methodologies facilitate on-vehicle decision-making processes based on locally accessible information, thereby addressing latency requirements in safety-critical systems such as collision avoidance, autonomous driving, and real-time traffic management. Faster and more adaptive responses are especially beneficial in V2X environments, where delays in data processing could compromise the safety and efficiency of the system.


\textit{Challenges of Distributed AI for V2X.}
The benefits of distributed AI are well understood, but can only be realized if some significant challenges, especially concerning the learning phase of such solutions, are addressed \cite{Alhammadi8845070}. One notable challenge is the heterogeneity of devices and data. Computing, communication, storage, and power resources, as well as the type of information collected (e.g., data types, data distribution), can vary significantly between vehicles and  \acp{RSU}. This heterogeneity makes it difficult to ensure effective cooperation between devices, potentially leading to sub-optimal performance. 

Although significant communication gains are expected through the limitation of data exchanges, distributed training processes often require constant model update transmissions from computing devices. Such model updates could consist of complex neural networks comprising millions or even billions of parameters. This implies a significant communication overhead, making the uplink transmissions particularly challenging. On the one hand, from the computing device perspective, transmitting a large volume of model updates periodically, imposes notable communication and energy demands. On the other hand, the reliability of distributed AI approaches can be significantly affected in dynamic network environments with time-varying conditions, such as fluctuating channel quality or traffic congestion, leading to potential delays or system failures. Security is another major concern, as risks such as model poisoning, adversarial attacks, and model theft can compromise the adoption of distributed AI solutions in V2X scenarios \cite{9644368}.

In addition, whatever the specific AI deployment, whether distributed or centralized, and the targeted usage, AI-related features need to be understood. Sophisticated solutions equipped with complex AI systems exist in the CCAM realm that analyze huge amounts of sensor/network data. However, the ‘black box’ nature of traditional AI could be a barrier of trust between  AI-based CCAM solutions and their potential users.

\textit{Addressing Challenges of Distributed AI for V2X.} In light of the aforementioned challenges, the implementation of dynamic and real-time optimization methods that take into account device, data and network dynamics has the potential to enhance distributed AI's reliability. Real-time traffic analysis and forecasting could effectively assist proactive decision-making and adaptation actions related to device selection, resource allocation, model and data exchange frequency, as well as the compression level of the transmitted information (e.g., model updates). A number of techniques have been developed for the latter, including AI model pruning, quantization, knowledge distillation, and low-rank factorization \cite{Dantas2024}. These techniques have the potential to not only reduce model size, but also speed up inference times for resource-constrained devices. 
\begin{figure}[h!]
    \centering
    \begin{subfigure}{0.9\linewidth}
        \centering
        \includegraphics[width=\linewidth]{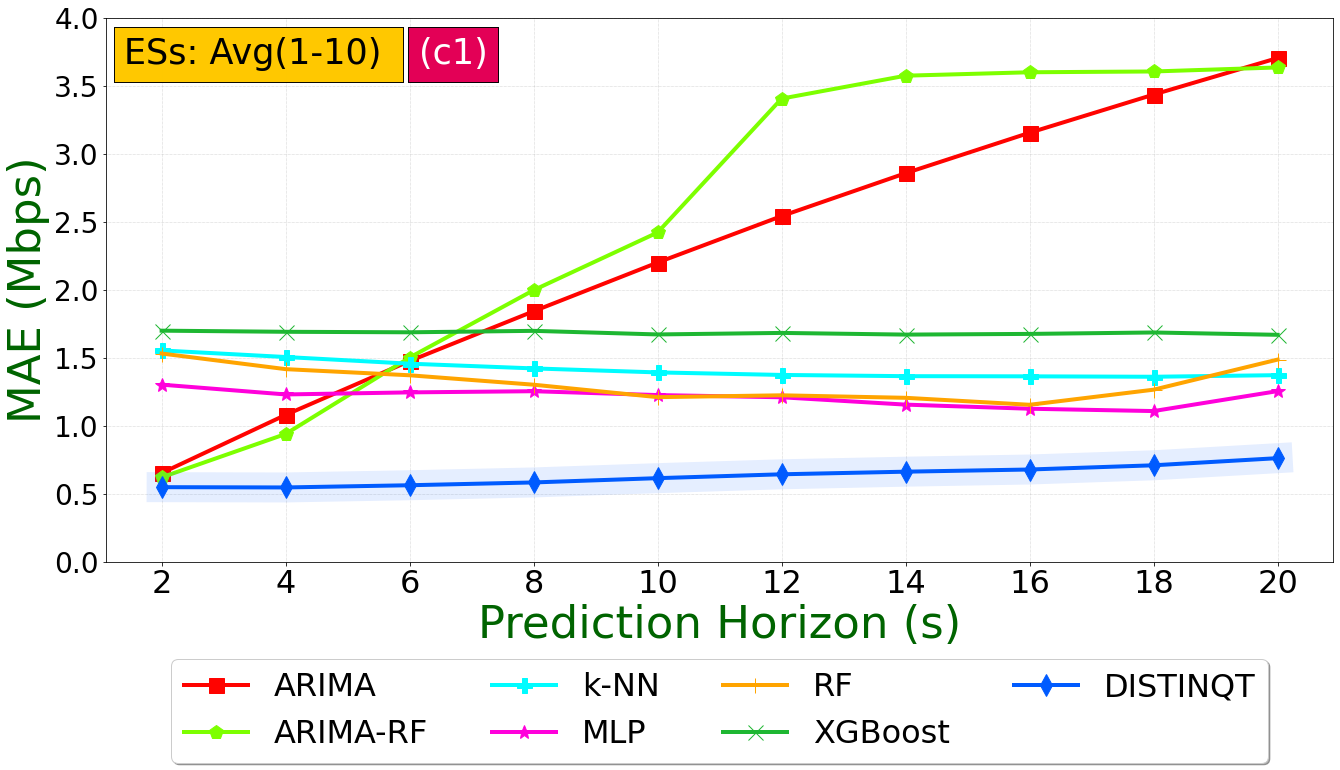}
        \caption{}
        \label{mae_all_baselines_4a}
    \end{subfigure}
    \vspace{8pt} 
    \begin{subfigure}{0.9\linewidth}
        \centering
        \includegraphics[width=\linewidth]{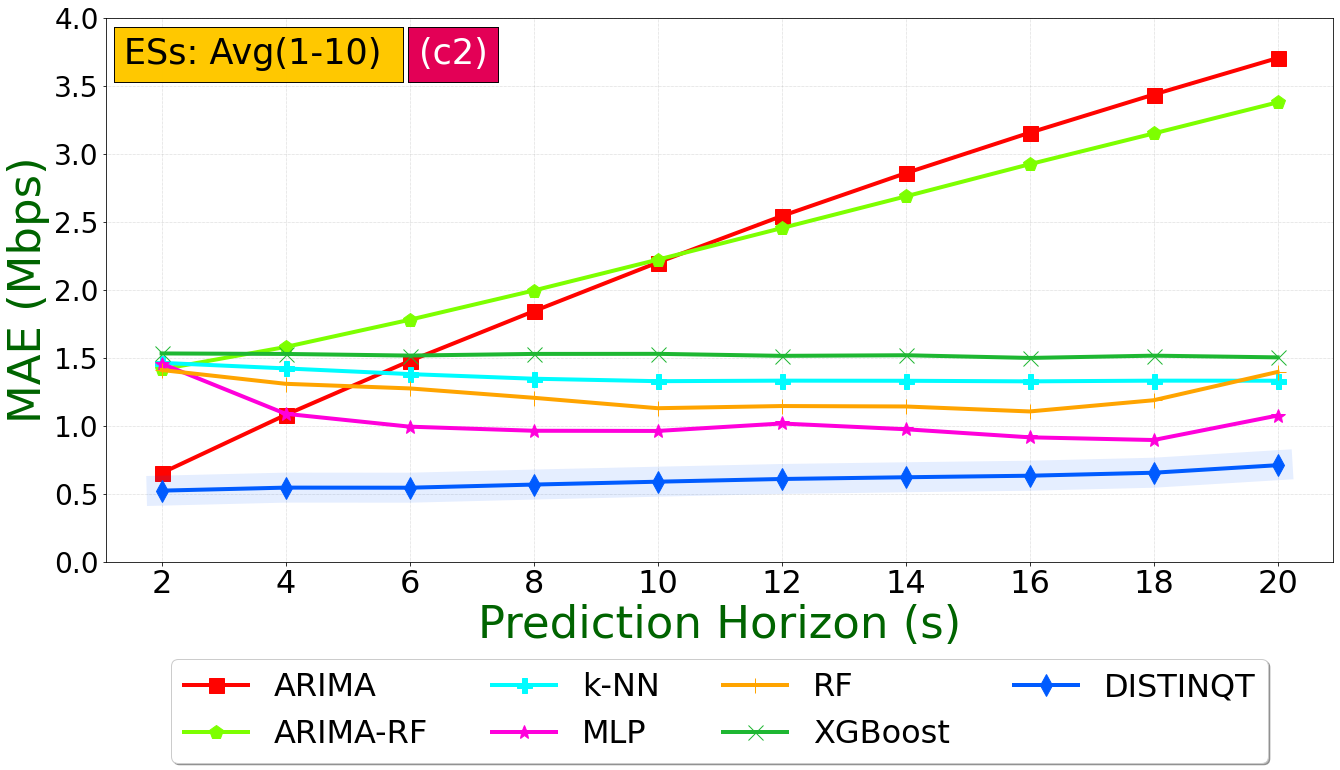}
        \caption{}
        \label{mae_all_baselines_4b}
    \end{subfigure}
    
    \caption{QoS prediction performance of DISTINQT compared to six centralized state-of-the-art methods across two configurations.}
    \vspace{-5pt}
\end{figure}
In V2X scenarios requiring data exchanges for cooperative functions or complex processing beyond a vehicle’s resource availability, protecting privacy and preventing malicious interference are critical concerns. Techniques such as differential privacy, encryption, and privacy-aware distributed AI using autoencoders, as demonstrated in \cite{koursioumpas2024distinqt}, have shown promising results in this direction. The work presented in \cite{koursioumpas2024distinqt}, referred to as DISTINQT, is a distributed PQoS solution that achieves privacy preservation by encoding raw data into compressed, non-reversible latent representations prior to transmission. This method ensures that sensitive information remains protected throughout the communication process. 

To validate the benefits of distributed AI for CCAM use cases, DISTINQT has been used on the same PQoS application of uplink throughput prediction in \ac{ToD} in Fig.~\ref{fig:PQoS}. Figures \ref{mae_all_baselines_4a} and \ref{mae_all_baselines_4b} illustrate the performance of DISTINQT compared to six state-of-the-art centralized baseline solutions for different prediction time horizons. The comparison is conducted under two distinct configuration setups: \textbf{c1}, where computations are distributed between \acp{UE} and \acp{BS}, and \textbf{c2}, which extends the distribution to also include \ac{MEC} servers. DISTINQT outperforms all baseline solutions in all selected prediction time horizons and under both configurations, retaining a stable performance as the prediction time horizon increases. 

\begin{figure}[h!]
    \centering
   \includegraphics[width=0.9\linewidth]{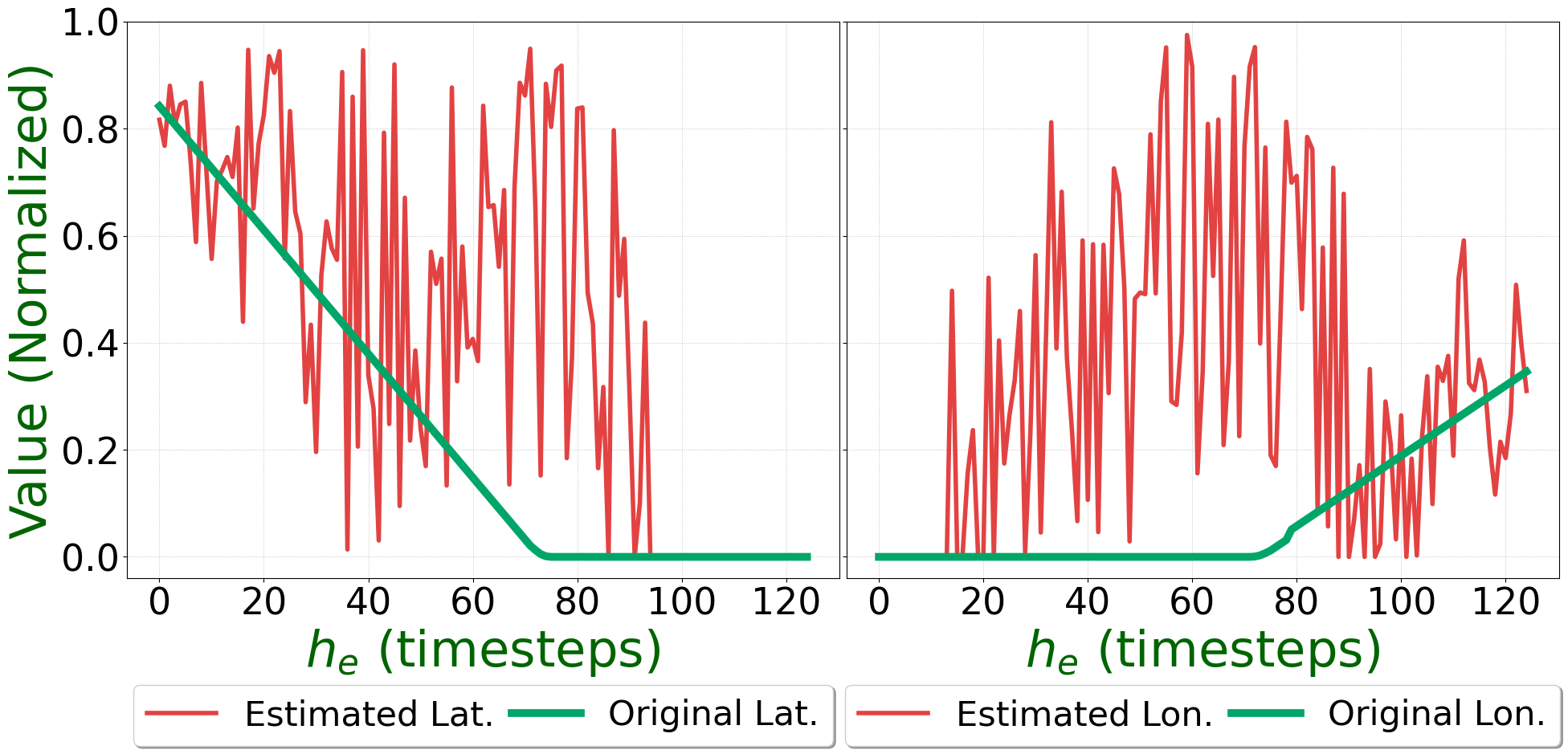}
    \caption{Original vs Estimated Sensitive Input Sequences.}
    \label{privacy_exp}
    \vspace{-8pt}
\end{figure}

Apart from prediction performance, the effectiveness of DISTINQT in preserving privacy is evaluated by examining an optimization problem with the objective of approximating the initial raw input data. Figure \ref{privacy_exp} depicts a comparison between the original and estimated sensitive data (i.e. the location history of the ToD), indicating a low similarity percentage equal to $20.26\%$ and $17.84\%$ for the latitude and longitude, respectively.
Lastly, strengthening security through advanced techniques like adversarial training and anomaly detection is crucial to safeguarding distributed AI models against adversarial threats. 

Finally, \ac{XAI}~\cite{xai} techniques need to be developed and adapted for the CCAM context to make decisions and predictions of AI systems understandable and transparent to humans.


\subsubsection{AI-as-a-Service \& Agentic AI}
\ac{AIaaS} corresponds to a paradigm shift where AI does not need to be created by each entity requiring AI, but where a 6G network offers pre-built AI models that both the network itself and vertical applications, such as CCAM, can readily access through \acp{API}~\cite{syed2025artificial}.   \textcolor{black}{In so doing, this paradigm targets both operational and service intelligence (discussed in detail in Section IV.D.\ref{sec:EOS}). On the one hand, AI models can support complex multi-objective decision making in allocation, orchestration,
and management of communication, computation,
sensing, and AI resources across the CCAM
ecosystem, by exploiting 
real-time operational data related to the state of communication
networks, edge/cloud resources, vehicles,
roadside infrastructure, energy resources, and AI workloads. On the other hand, AI models transforming service-related data into knowledge  may enable CCAM services in a more  personalized manner.}

AIaaS has been identified by the Hexa-X II project\footnote{\url{https://hexa-x-ii.eu/}} as one of the major enablers for 6G systems.  

\begin{figure}[h!]
    \centering
   \includegraphics[width=0.9\linewidth]{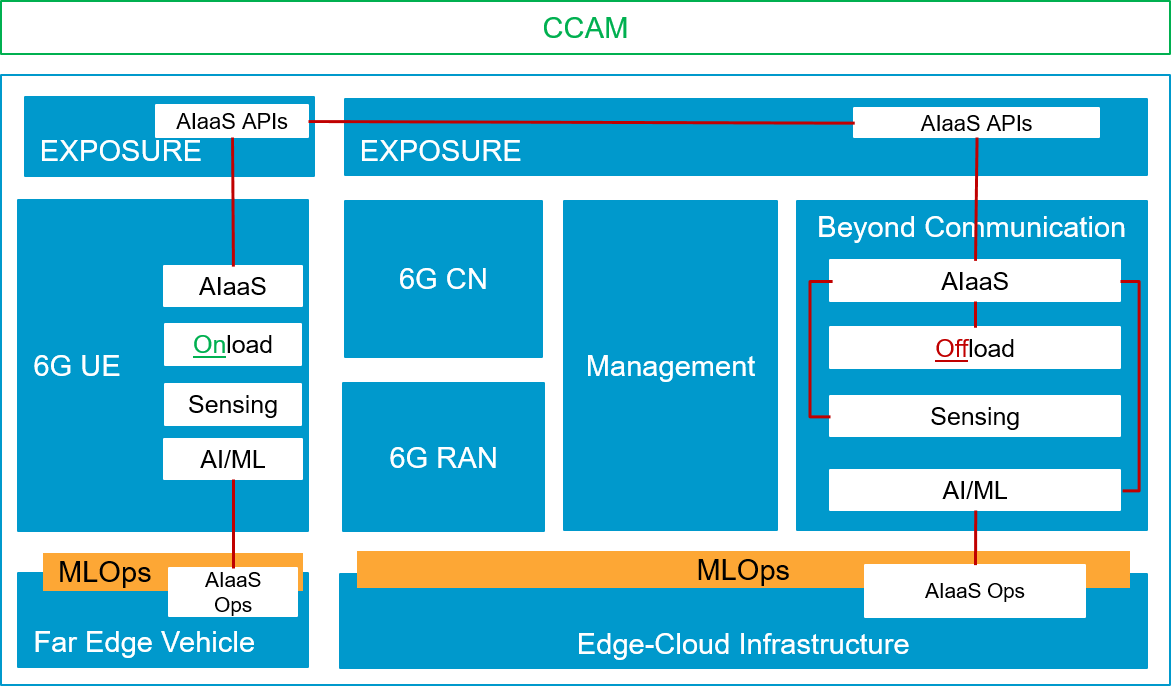}
    \caption{
    AIaaS for CCAM (modified from\cite{AIaas_Ericsson}).}
    \label{aiaas_ccam}
\end{figure}

An illustration of \ac{AIaaS} for 
CCAM is depicted in Fig.~\ref{aiaas_ccam}. 
A traditional interpretation corresponds to a set of \acp{API} (offloading, sensing, \ac{ML} operations) in a 6G network used to discover available compute capabilities in heterogeneous edge-cloud infrastructures and subsequently to {\em off}-load AI/ML entities accordingly. However, CCAM extends this interpretation, where these APIs are used to discover compute and sensing capabilities also in far edge vehicles, where AI/ML entities could be {\em on}-loaded for inference or training. 

\textit{Challenges of AIaaS:}
\ac{AIaaS} for CCAM acts as an enabler to involve far-edge nodes (vehicles, \acp{RSU}, \ac{IoT} devices) in AI creation or AI transfer between heterogeneous and decentralized AI producers and consumers. \ac{AIaaS} for 6G CCAM however faces several challenges.

\begin{itemize}

    \item Real-time coordination exposure: 
    CCAM use cases such as cooperative maneuvering, platooning, or intersection management require deterministic AI inference and tightly synchronized decision-making across multiple vehicles and infrastructure nodes. AIaaS must therefore expose timing guarantees, execution deadlines, and coordination primitives that ensure bounded latency when off-loading or on-loading AI/ML entities.

    \item Safety-compliant model exposure: unlike generic verticals, CCAM requires that AI models meet functional safety and regulatory constraints. AIaaS must expose metadata on model validity, certification status, operating domain, and safety envelopes, enabling vehicles and RSUs to verify that shared perception or prediction models can be safely integrated into automated driving functions.

    \item Collective perception and sensor alignment exposure:
    CCAM depends on \ac{CP} across vehicles equipped with heterogeneous sensing suites (LiDARs, radars, cameras, GNSS). AIaaS must expose calibration parameters, sensing uncertainties, and spatiotemporal alignment information to enable consistent fusion of perception outputs across far-edge nodes.

    \item Mobility-aware orchestration exposure: vehicles continuously move across coverage areas, administrative domains, and dynamic traffic conditions. AIaaS must provide APIs for mobility-aware AI orchestration, including predictive placement or migration of AI/ML entities, adaptive switching between on-board and off-board inference, and maintaining performance under varying network load and connectivity.

    \item Resilience and fail-operational exposure: CCAM requires that AI functions remain operational despite network disruptions, compute overload, or faulty model updates. AIaaS must therefore expose fail-over capabilities, such as automatic fallback to validated models, rerouting inference to alternative edge nodes, and maintaining continuity of perception and prediction services.\\

\end{itemize}

\textit{Addressing the challenges of AIaaS:}
Addressing the above challenges requires advances in orchestration, safety assurance, \ac{CP}, mobility handling, and system resilience. Ensuring real-time coordination among vehicles and \acp{RSU} demands orchestration mechanisms that can maintain deterministic inference latency and synchronized manoeuvre execution; constraint-driven approaches such as model predictive control have demonstrated effective coordination of automated vehicles at intersections~\cite{Mihaly2020MPCIntersection}. Meeting safety and regulatory requirements further requires model metadata exposure, certification tracking, and runtime monitoring so that CCAM functions rely only on validated and trustworthy AI components for AI-driven automated systems~\cite{PerezCerrolaza2024AISafety}. \ac{CP} across heterogeneous sensors necessitates exposing calibration parameters, uncertainty characterization, and alignment mechanisms to ensure robust fusion of perception streams from multiple vehicles and infrastructure nodes; techniques building on robust model predictive control of complex multi-vehicle geometries illustrate how uncertainty handling and coordinated perception can be strengthened in such settings~\cite{Farkas2023MPCRoundabouts}. Because vehicles and workloads are mobile, AI functions must be dynamically migrated \textcolor{black}{(across vehicles, roadside infrastructure, and
cloud/edge servers)} or adapted based on predicted trajectories, \textcolor{black}{accuracy requirements,} resource availability, and radio conditions, motivating mobility-aware orchestration mechanisms at the vehicular edge. 
\textcolor{black}{The aforementioned objectives in the migration decisions should be also balanced with broader
sustainability goals, including energy efficiency, emissions
reduction, and carbon footprint minimization \cite{10458416Koursioumpas, panagea2026}.}
Finally, CCAM systems require fail-operational behaviour even under disruptions, motivating redundancy, safe fallback, and validated model reversion strategies for automated driving~\cite{vomDorff2020FailSafe}. Collectively, these approaches establish a foundation for a 6G AIaaS platform capable of supporting safe, predictable, and cooperative intelligence across far-edge vehicular and roadside nodes.

\subsubsection{Emerging Transversal Approaches for AI Deployment in CCAM} \label{sec:EOS}

\textcolor{black}{The AI techniques discussed in the previous subsections address different but strongly interconnected aspects of future CCAM systems. Predictive QoS enables anticipation of network and service conditions; distributed AI allows intelligence to be deployed across vehicles, roadside infrastructure and edge/cloud resources; AI-as-a-Service and agentic AI provide mechanisms for discovering, exposing, orchestrating and consuming AI capabilities throughout the ecosystem. However, the effectiveness of these approaches depends on their ability to operate in a coordinated manner. In highly dynamic CCAM environments, AI deployment decisions cannot be taken solely based on communication, computation, or application requirements. They must also account for mobility patterns, traffic conditions, environmental context, energy availability, sustainability objectives, infrastructure status, and operational constraints. This creates the need for transversal intelligence frameworks capable of integrating heterogeneous information sources and coordinating AI-driven decisions across multiple system layers.\\
Such a framework can be provided by the combination of three emerging concepts in the field of AI that can be defined as follows in the context of CCAM:
\begin{itemize}
\item Environmental Intelligence (EI) - the capability to transform environmental observations, encompassing information such as weather conditions, road conditions, traffic density, air quality, visibility, moving obstacles, energy availability, and environmental hazards, and contextual data into actionable knowledge that improves perception, prediction, adaptation, and sustainability-aware decision making across CCAM systems \cite{11460159Lathabai}.
\item Operational Intelligence (OI) - already explored in fields like business \cite{OI_business} and industrial processes \cite{OI_industry}, it can be defined as the capability to transform real-time operational data related to the state of communication networks, edge/cloud resources, vehicles, roadside infrastructure, energy resources, and AI workloads, into decisions that optimize the allocation, orchestration, and management of communication, computation, sensing, and AI resources across the CCAM ecosystem.
\item Service Intelligence (SI) - the capability to transform service-related data into knowledge that enables prediction, optimization, and assurance of service performance, reliability, safety, and user experience across CCAM applications \cite{SI_reading109853}.
\end{itemize}
The combination of these three concepts, leading to the definition of Environmental, Operational and Service Intelligence (EOS-I) can provide a unified intelligence layer spanning environmental sensing, service management, operational monitoring, communication networks, and decision-support mechanisms. 
EOS-I can enhance CCAM decision-making in three main ways. First, it can support environmental-aware perception, where vehicles and infrastructure jointly interpret road, weather, traffic, and environmental conditions for improved coordination, incident detection, and infrastructure adaptation. Second, it can enable predictive and adaptive V2X operation. Third, it can support impact assessment by evaluating how V2X services affect energy consumption, emissions, infrastructure utilization, traffic flow, and urban sustainability. This direction is aligned with the CCAM Partnership’s Strategic Research and Innovation Agenda, which identifies sustainability, traffic efficiency, safety, and environmental impact as core dimensions for future CCAM systems \cite{CCAMPartnership2022SRIA}.}

\subsection{Virtualization}


 By abstracting the capabilities of potentially heterogeneous physical components, virtualization provides high flexibility in order to adapt to
a variety of different scenarios and requirements.  
Two relevant instantiations for CCAM services in the application and network domains are presented in the following. 



\subsubsection{Digital Twins}

By building dynamic digital models of the physical world and establishing bidirectional communication between the real and cyber worlds, \acp{DT} can simulate, predict, and analyze the behavior of their real-world counterparts,
thus facilitating and improving functionalities such as real-time monitoring, predictive maintenance, optimized design, and proactive decision making in several sectors. 
CCAM stands as one of the most prominent application areas for \acp{DT} \cite{DT-CAV}, \cite{mobilityDT}. 

\begin{figure}[t]
\centering
\includegraphics[width=0.95\columnwidth,draft=false]{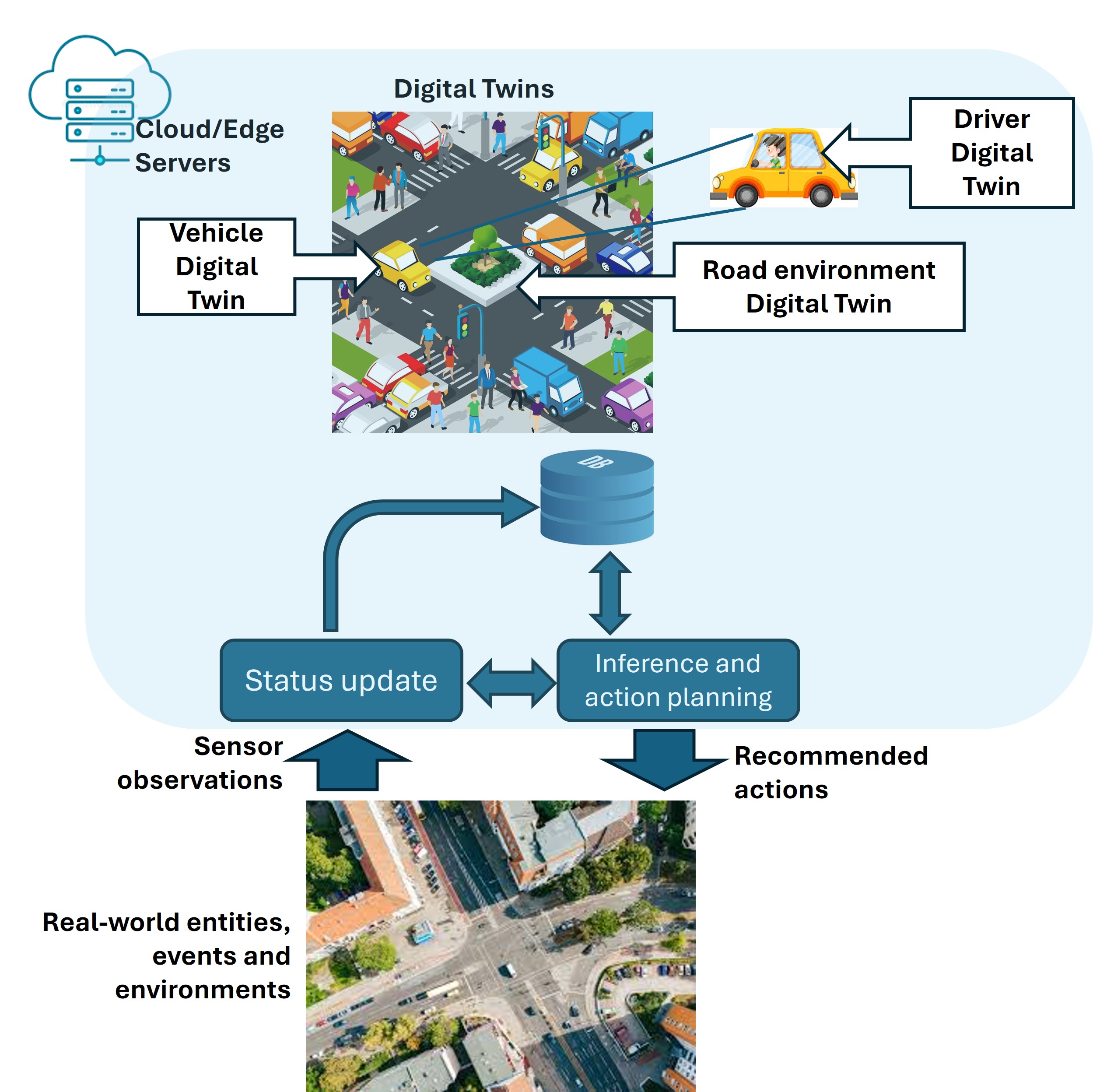}
\caption{DTs in CCAM scenarios (adapted from\cite{aecc24}). 
}
\label{fig:AECC-DT}
\end{figure}

In CCAM systems, \acp{DT} can be instantiated at different levels of granularity: from modeling road infrastructure to representing individual vehicles \cite{aecc} and drivers 
\cite{VNC24}, 
\cite{ma2024driver}. An example is shown in Fig.~\ref{fig:AECC-DT}; the lower portion depicts the physical road environment, including, e.g., vehicles, drivers, vulnerable road users, road infrastructure; the upper portion illustrates the corresponding \acp{DT}. Such digital counterparts are continuously updated using sensor data that reflect their real-time status. In addition, \acp{DT} can store historical data for further analysis. The collected and inferred data support decision-making processes, which in turn generate  recommended actions that influence and optimize the behavior of the physical entities. Numerous studies have demonstrated the effectiveness of vehicle \acp{DT} interacting with other digital entities in enhancing CCAM performance: from \ac{ADAS} \cite{wang2020digital} and \ac{CP} \cite{li2023het} to navigation \cite{wang2024smart} and \ac{ToD} \cite{Lentisco2024}. 

\textit{Challenges of DTs in V2X.} 
Despite their potential, several challenges hinder large-scale deployment of \acp{DT} in CCAM contexts 
\cite{mobilityDT}, \cite{haverkamp2025digital}. 
Sensing and modeling accuracy remain major concerns. Building a precise digital replica of the physical environment requires diverse sensors (LiDARs, radars, cameras, positioning and vehicle dynamics sensors, etc.) and advanced modeling. Errors from noise, occlusions, misalignment, or latency can cause divergence between the physical and digital worlds, compromising safety-critical functions like collision avoidance. Accurate, real-time modeling of wireless channel dynamics also remains difficult due to complex, non-ideal environments \cite{DTOCM}.

Interoperability and standardization are crucial. DTs integrate data from vehicles, infrastructure, networks, and cloud systems developed by different vendors. The absence of unified data models, interfaces, and protocols limits cross-domain collaboration and large-scale integration.
Physical-digital communication is another concern. In particular, communication latency and reliability are critical for real-time \ac{DT} applications such as cooperative driving or hazard response. Achieving low-latency, high-reliability links demands robust 5G/\ac{URLLC} or future 6G networks, efficient edge computing, and resilient synchronization between physical and digital systems.

Computational scalability is another issue. High-fidelity \acp{DT} require significant processing power, memory, and bandwidth. Balancing model accuracy with computational feasibility is essential for real-time responsiveness.

Finally, security and privacy are foundational to safety and public trust \cite{DTsec}, \cite{DTsec2}. \acp{DT} are vulnerable to data tampering, desynchronization, and adversarial attacks, while continuous connectivity exposes vehicles to eavesdropping and denial-of-service threats. Moreover, DT-based services often process sensitive behavioral and location data, raising privacy and regulatory concerns. Ensuring cybersecurity, data anonymization, and secure governance is therefore critical.

\textit{Addressing challenges of DTs for V2X.} 
Research is advancing several strategies to overcome DT challenges in CCAM systems.

Improving sensing and modeling accuracy relies heavily on \ac{ML} and multimodal data fusion to enhance robustness against noise and missing data. High-fidelity simulation frameworks integrate realistic sensor \cite{cazzella2024multi}, environment, and communication models to minimize the sim-to-real gap \cite{brunner2025r}, improving DT reliability in autonomous driving and intersection management 
\cite{ali2025b}.
\textcolor{black}{Integrating \ac{LLM} technology into \acp{DT} will further help realize EOS-I, in particular the environmental intelligence, by including not only simple perception of the physical environment,
but also a comprehensive understanding and dynamic response
to environment changes, channel characteristics, etc. \cite{yu2025channelgpt}.}

Reducing latency is addressed through edge computing \cite{wang2024digital}, \ac{URLLC} \cite{van2022edge}, and hierarchical (sensor–fog–cloud) architectures \cite{haverkamp2025digital} that distribute computation closer to data sources, ensuring real-time synchronization and scalability. 

Interoperability and standardization efforts focus on developing shared APIs, data formats, and update protocols to enable cross-vendor integration. However, despite growing global interest and open-source collaboration initiatives \cite{DTconsortium}, there is still no consensus on DT implementation \cite{DTarchi}. Rather than building DTs from scratch, leveraging existing open-source DT frameworks can accelerate adoption while avoiding redundant development 
\cite{conde2023collaboration}, \cite{OMA}.

Enhancing security, privacy, and trust involves integrating blockchain for data integrity, federated learning for privacy-preserving AI, and advanced cryptographic tools (e.g., homomorphic encryption, secure multiparty computation, differential privacy). Context-aware, zero-trust architectures, blockchain-based identity management, and AI-driven intrusion detection are also being explored to safeguard distributed DTs. Lightweight, scalable trust mechanisms remain a key research priority \cite{DTsec}, \cite{DTsec2}.

\subsubsection{Software-Defined Internet of Vehicles}

Software-defined Internet of vehicles (SDIoV)\acused{SDIoV} offers a new way to manage the growing complexity of connected and automated vehicle networks. Grounded in \ac{SDN}, SDIoV separates network control functions from the physical hardware, enabling more flexible, programmable, and scalable communication systems. This approach is particularly important for CCAM, where low latency, high reliability, and coordination between diverse devices are essential~\cite{Jittrapirom2017}.

By introducing centralized and programmable control, SDIoV supports key functionalities such as traffic coordination, environmental awareness, and safety-related decision-making. The SDN principle allows the network to optimize data transmission and routing based on a global view. When combined with edge or fog computing, part of the data processing happens closer to the vehicles, which reduces communication delay and improves responsiveness. For example, the Amsterdam pilot project on edge-enabled autonomous mobility showed how edge-enabled SDIoV can significantly improve traffic safety and efficiency by enabling direct communication between vehicles and roadside infrastructure through 5G and nearby edge servers~\cite{Sandu2021}.

The SDIoV framework typically includes several layers that work together for real-time vehicular communication. At a high level, it involves a central controller, smart network switches, roadside infrastructure, and in-vehicle communication units. These elements are organized into logical layers: the application and control layers handle decisions and service logic; the data plane layers manage network connectivity through base stations, roadside units, and vehicles; and the knowledge layer collects environmental and traffic data to support network optimization~\cite{Pangbourne2020}. 

\begin{figure}[t]
\centering
\includegraphics[width=1\columnwidth]{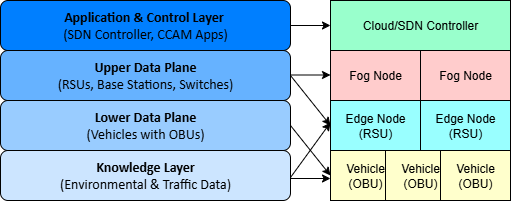}
\caption{SDIoV architecture.}
\label{sdviot}
\end{figure}

Figure~\ref{sdviot} illustrates the SDIoV paradigm by considering both its architectural layers and their concrete realization in CCAM use cases: the left part outlines the layered SDIoV architecture, comprising the application and control layer, upper and lower data planes, and the knowledge layer. These abstract layers directly map to the operational components shown on the right, where the application and control layer is embodied by the cloud-based SDN controller, the upper data plane corresponds to fog nodes and \acp{RSU} acting as edge nodes, and the lower data plane is realized by vehicles equipped with \acp{OBU}. Meanwhile, the knowledge layer is supported by sensed environmental and traffic data continuously collected from vehicles and infrastructure. This mapping shows how SDIoV enables low-latency decision-making and coordinated control across vehicles and infrastructure.

Two main extensions of this architecture are commonly discussed. Edge-enabled SDIoV allows vehicles and nearby infrastructure to make quick local decisions for routing or hazard detection. Fog-enabled SDIoV adds an intermediate layer between the edge and the cloud to improve coordination and load balancing across the system.

\textit{Challenges in SDIoV:} 
Deploying SDIoV in the context of CCAM faces several challenges. One key issue is network heterogeneity, since SDIoV must integrate different wireless technologies such as cellular and Wi-Fi-based solutions. High vehicle mobility also makes it difficult to maintain stable routing and control synchronization. Security and privacy are further concerns, as open and multi-stakeholder communication exposes the system to potential attacks. Finally, efficient allocation of communication and computing resources is critical to prevent congestion during peak traffic periods.

\textit{Addressing SDIoV challenges:} 
\ac{SDN}’s centralized intelligence enables the network to adapt to traffic density, link quality, or application needs in real time. By placing computation closer to vehicles and employing some of the previously mentioned technologies (e.g., \ac{PQoS}, \acp{DT}), edge-enabled \ac{SDIoV} reduces dependence on distant cloud resources and supports delay-sensitive functions such as collision avoidance. Fog nodes near \acp{RSU} can further reduce network load by filtering or aggregating data locally. Finally, SDIoV promotes collaborative resource sharing, where vehicles with spare capacity assist others nearby, creating a more resilient and adaptive network. These mechanisms make SDIoV a strong foundation for the future of connected and automated mobility.

\section {What we Have: Main Standardization and Regulatory Frameworks} \label{standard} 




The technological foundation for \ac{CCAM} is being laid within major standardization bodies, which are defining the protocols, interfaces, and architectures for core enabling technologies. While a significant portion of the initial focus has been on the physical and link layers of \ac{V2X} communication and the definition of messages for \ac{CCAM}, ongoing efforts are increasingly expanding to encompass the broader ecosystem, including the exploitation of novel spectrum (e.g., THz bands), utilizing \ac{RIS} for enhancing the propagation characteristics and coverage, integration of edge intelligence, and the  use of artificial intelligence to manage network and service complexity.

When looking at the \ac{V2X} communication technologies available today for CCAM, there is a distinction between classical cellular communications (i.e., downlink and uplink communications between devices and \acp{BS}) and direct communication between devices (often referred to as sidelink). Focusing on the \textit{classical cellular communications}, even if the general radio access features are exploited by connected cars without the need for specific modifications, new features for \ac{V2X} have been added, enhancing the architecture of LTE in 2017 \cite{chen2017vehicle}. In particular, new logical parts have been added to address the automotive scenario and enable service provision from external stakeholders. Also relevant for \ac{V2X} is the standardization effort on the topic of \ac{PQoS}, which allows applications to be notified about expected variations in the \ac{QoS} and thus react in advance \cite{9129065}. Another important aspect is the special focus on localization and \ac{ISAC} that will be further discussed hereafter.


Concerning \textit{direct communications}, two directions that have been taken in standardization are WiFi-based and  cellular-based sidelink. After more than two decades of activity in various groups, WiFi-based and cellular  technologies are standardized~\cite{anwar2019physical, NRV2XR16Tutorial}. The WiFi-based solution is today installed on 
some car models in Europe, while cellular sidelink appears at the moment the preferred choice in some parts of the world, such as the US and China. 

In terms of \textit{WiFi-based V2X}, IEEE 802.11p was defined in 2010 and the backward compatible update IEEE 802.11bd was published in 2023. This approach applied by IEEE to these standards is different compared to what is used by 3GPP for its generations. The various amendments of IEEE 802.11, in fact, at some point in time converge to a new update of a single comprehensive standard, which is retro-compatible with the previous ones. Retro-compatibility is mainly achieved through a common preamble and a basic set of features that are always available. The latest version, including both the mentioned amendments, is the IEEE 802.11-2024. Given the common framework with other WiFi-based systems, WiFi-based \ac{V2X} is capable of both unicast and broadcast transmissions. IEEE 802.11p had mainly incorporated the use of 10~MHz channels and the concept of communication without necessarily being authenticated as part of a network, which is called \ac{OCB}.  The IEEE 802.11bd then added \ac{NGV}, with a number of features targeting higher throughput, larger range, and better support for long packets in high mobility conditions \cite{Car2Car_11bd}. The IEEE 802.11bd also considers 20 MHz channels and the use of mmWave bands, and introduces specific procedures for fast ranging. 


Regarding the standardization of \textit{cellular sidelink},  LTE-V2X  was  specified in 2017 and 5G NR-V2X in 2020. 
5G NR-V2X sidelink (NR hereafter) reused several of the concepts existing for LTE-V2X sidelink (LTE hereafter) with a number of enhancements made throughout three releases. Going beyond the broadcast transmissions in LTE, new procedures were introduced in Rel-16 NR for supporting groupcast and unicast transmissions \cite{NRV2XR16Tutorial}, allowing communication between a pair of \acp{UE} or between a \ac{UE} and a group of \acp{UE}. Resource allocation enhancements were introduced in Rel-17 including inter-UE coordination, allowing a UE to provide resource recommendations to another UE to support enhanced reliability and reduced latency. Two new spectrum options were considered in Rel-18 for NR, namely, unlicensed spectrum and millimeter wave (FR2) licensed spectrum, targeting an increase of sidelink data rate and an expansion of the applicability of sidelink to commercial use cases. Channel access mechanisms and physical channel design for unlicensed spectrum were specified in Rel-18. Regarding FR2 licensed spectrum, the support of sidelink beam management, i.e., beam pairing and beam maintenance, was studied, but only considering unicast communication.

Independently from the access technology, standards also define the messages that need to be exchanged through direct communications, including broadcast, groupcast and unicast. The first broadcast messages were defined to share vehicle's own information, then updated to also share what is sensed by the vehicles and road-side units, and later to coordinate the manoeuvres. Examples of these types of messages are, in the European \ac{C-ITS} standards under development within ETSI, respectively, \acp{CAM}~\cite{festagCAM}, \acp{CPM}, and \acp{MCM}~\cite{thandavarayan2020cooperative}. Activity is also undergoing within ETSI for the definition of rules for radio resource management towards an efficient and coordinated use of the multiple available channels \cite{10467184}.

Two other specific topics for CCAM that have also been discussed in the standardization bodies include sidelink positioning and ISAC. \textit{Sidelink positioning} was studied and specified within 3GPP Rel-18 as part of the expanded and improved NR positioning. The transmission of reference signals for sidelink positioning was specified for shared resource pools and dedicated resource pools. In the latter, a resource pool is configured just for the transmission of positioning reference signals in the sidelink. For improved resource efficiency, the transmission of positioning reference signal sent by Rel-18 UEs can share the resources with Rel-16/17/18 sidelink communication in a shared resource pool. A Rel-18 UE can transmit a positioning reference signal which can be multiplexed in time with a data transmission. In this case, backward compatibility with legacy Rel-16/17 UEs needs to be ensured, i.e., legacy Rel-16/17 UEs need to be aware of the resources reserved for sidelink positioning reference signals.



In addition, \textit{standardization of ISAC in cellular systems} has recently gained particular interest. 
Work at 3GPP started with a Rel-19 study on ISAC use cases, which identified several use cases and performance requirements for smart transportation \cite{3GPP22837}, including intrusion detection on a highway or a railway, sensing-assisted automotive maneuvering and navigation, sensing at crossroads, accurate sensing for automotive maneuvering and navigation service, vehicle sensing for advanced driver-assistance systems, and blind spot detection. 
This work was followed by a 3GPP Rel-19 study which focused on defining channel modelling aspects to support the detection and/or tracking of objects including: UAVs, humans indoors and outdoors, automotive vehicles, automated guided vehicles, and objects creating hazards on roads or railways. 
As a result of this study, the 3GPP technical report on channel models for frequencies from 0.5 to 100 GHz \cite{3GPP38901} has been extended to include the channel modelling framework developed for ISAC in Rel-19. 
For Rel-20, 3GPP has recently approved two ISAC-related studies: a study on ISAC for 5G NR \cite{3GPPRP-251861} and a study on 6G radio access which includes an objective on sensing \cite{3GPPRP-251881}. The focus of the first study is on the performance evaluation of gNB-based mono-static sensing for the detection and tracking of UAVs, with the existing downlink NR waveform and reference signals  as baseline. 
In contrast, the \hbox{Rel-20} 6G study on sensing does not need to restrict itself to 5G \ac{PHY} functions and procedures or to the scenario considered in the \hbox{Rel-20} 5G study. This 6G study on sensing will consider use cases and requirements resulting from another study on 6G scenarios and requirements. Aspects regarding the integration of sensing with communication services are also part of the scope of the 6G study on sensing.

In terms of \ac{RIS} and THz as enablers for \ac{CCAM}, standardization efforts are still in their early stages, but the growing interest of bodies such as ETSI and 3GPP is catalyzing foundational studies on the integration of these technologies into future networks. 
\textit{\ac{RIS}}, in particular,  have recently emerged as a new subject for standardization, aiming to bring programmability to the wireless environment, which is of particular interest for highly mobile, cluttered CCAM propagation environments. The lead body is the ETSI ISG RIS, which is currently in the process of building a foundational framework for the technology \cite{ETSIIISGRIS}. Its work involves defining use cases and KPIs, architectural principles for RIS control (e.g., centralized vs. distributed), channel models for RIS, and implementation and practical considerations of RIS deployment. \ac{3GPP} has also initiated studies related to RIS (under the term "smart repeaters") within the context of its Rel-18 and Rel-19 work items. The focus is on assessing the feasibility and benefits of integrating RIS into the cellular network to enhance coverage, a study that directly informs the potential future standardization of RIS for improving \ac{C-V2X} link reliability, particularly in challenging radio conditions \cite{3GPP22837}.

Also at an early stage from a standardization point of view are \textit{THz communications}, which are expected to unlock extreme data rates for future CCAM, in particular aiding high throughput functionalities such as sensor data exchange. In IEEE, the 802.15.3 Task Group 3d (TG 3d) has produced IEEE 802.15.3d-2017 standard, which specifies a physical layer for switched point-to-point links operating in bands above 252 GHz \cite{IEEE802.15.3d}. While dealing with point-to-point links only, this standard represents one of the first formal specifications for THz communications, establishing a basis for future technology development. 
Going beyond point-to-point links, the work addressing THz in the scope of future mobile networks is carried out within the ETSI Industry Specification Group on THz (ISG THz). Its work is focusing on identifying use cases relevant to both fixed and mobile links, developing channel models (including for mobile scenarios), and defining RF hardware models and initial system requirements for THz systems \cite{ETSIISGTHz}. This work aims to define the necessary technological base for future incorporation of THz bands into broader communication standards.

More advanced is instead the standardization of \textit{edge computing} architectures, which is critical for defining how low-latency CCAM applications are hosted and managed at the network edge. The ETSI \ac{ISG} on \ac{MEC} is the central standardization body for this effort. Particularly relevant for CCAM is the specification of the \ac{V2X} Information Service \ac{API} (ETSI GS MEC 030) \cite{ETSIMEC030}. This API standardizes how vehicles, \acp{RSU}, and CCAM application servers can retrieve contextual information (e.g., a localized dynamic map, list of nearby entities) from a MEC platform, enabling a standardized approach to centralized coordination. The 3GPP complements this by standardizing the network procedures that enable MEC. Specifications for the 5G System Architecture (3GPP TS 23.501) define the mechanisms for traffic steering and session management that ensure CCAM data flows are routed to a local functions co-located with a MEC host, thereby creating a standardized end-to-end path for edge computing \cite{3GPP23501}.

Finally, \textit{\ac{AI}} technology is also a topic in many standardization bodies, primarily for network automation and optimization, with implications for CCAM service reliability. Within the 3GPP, the most concrete effort is the standardization of the \ac{NWDAF} as a core network function (3GPP TS 23.288). The NWDAF provides a standardized framework for collecting network data, training ML models, and producing fanalytics that can be consumed by other network functions or authorized applications \cite{3GPP23288}. For CCAM, this enables standardized, AI-driven predictions of network performance. Beyond the NWDAF, 3GPP has completed study items on AI for the air interface (e.g., for channel state prediction), with the results feeding into ongoing normative work. 
With similar objectives,  initiatives such as AI-RAN\footnote{AI-RAN - \url{https://ai-ran.org/}} are targeting AI-native orchestration of multiple intelligence layers into one continuum-ready framework. 

In parallel, ETSI addresses AI from a system architecture and operational perspective. ETSI ISG ENI is standardizing an AI-based architecture for context-aware network policy management, which can be applied to dynamically manage resources for CCAM services \cite{ETSIENI}. Furthermore, the ETSI ISG MEC is working on standards for deploying and orchestrating AI workloads on edge platforms, which is pivotal for standardizing how distributed intelligence for collaborative perception and other CCAM applications is instantiated and managed at the network edge.

\section{Conclusions and Future Research Directions} \label{concl}

This paper has provided a systematic review and analysis of the use case,  enabling technologies,  future directions, and standardization landscape for \ac{CCAM}. The main impediment to mass \ac{CCAM} deployment is the inability to integrate non-connected road users, a challenge that technologies like \ac{CP} are poised to address by enabling a distributed detection of cooperative and non-cooperative users and objects. While current standardization (3GPP C-V2X, IEEE 802.11bd) provides a necessary foundation for connectivity, it is insufficient for high-precision \ac{CP} and consequently for enabling high-reliability, low-latency, and integrated sensing demands of advanced CCAM use cases like intersection management, \ac{VRU} protection, and teleoperated driving.

Our analysis indicates that the key enablers for overcoming these limitations will converge with the development of 6G systems. Technologies such as \ac{ISAC} with addition of new, larger frequency bands, will enhance the sensing capability and reliability. Pervasive artificial intelligence, from \ac{PQoS} to distributed learning frameworks, will be critical for real-time decision-making and resource management. Furthermore, architectural paradigms like \acp{DT} and the \ac{SDIoV} will provide the necessary flexibility and control for system-wide optimization.

Significant research challenges remain to enable the full CCAM vision. Future work will need to address the following directions:
\begin{itemize}
    \item \textit{Integrated co-design of communication, sensing, and AI:} Moving beyond isolated optimizations to develop cross-domain, cross-layer frameworks that jointly manage communication, sensing, computation, and intelligence.
    \item \textit{Standardization:} Creating standardized and scalable CCAM architectures and semantic data models to ensure interoperability across vendors and facilitate large-scale deployment, feeding results into ongoing standardization efforts.
    \item \textit{Robust Security:} Developing lightweight, resilient security protocols to protect against adversarial attacks on AI models, V2X links, and sensor data, which is paramount for functional safety.
    \item \textit{Validation:} Establishing open, large-scale experimental testbeds for validating the performance of integrated CCAM technologies under realistic conditions (e.g., building on top of the ongoing efforts such as  SIMON\footnote{\url{https://www.bmv.de/SharedDocs/DE/Artikel/DG/mfund-projekte/simon.html}} and CARRISMA\footnote{\url{https://www.thi.de/forschung/carissma/labore/car2x-labor/}}).
\end{itemize}

In conclusion, while full CCAM deployment is not yet a reality, this paper aims at defining the trajectory towards it. The development of 6G technologies presents a concrete opportunity to address the existing technical gaps to enable CCAM. However, the focus will need to shift to integrated system design, rigorous validation, and standardization, in order to transition from research prototypes into safe, efficient, and ubiquitous cooperative mobility systems.


\section*{Acknowledgement}
This paper is an outcome of the work performed within one6G Work Item on Cooperative, Connected, and Automated Mobility (CCAM) 
(\url{https://one6g.org/working-groups/working-group-2/}).

\bibliography{refer.bib}
\bibliographystyle{ieeetr}

\end{document}